\documentclass[jair,twoside,11pt]{article}
\usepackage{jair, rawfonts}
\usepackage[backend=biber, style = apa, natbib = true]{biblatex}
\usepackage{url}
\usepackage[normalem]{ulem}

\usepackage[utf8]{inputenc}
\usepackage{hyperref}
\usepackage[paper=a4paper,headsep=0.5in,headheight=0.5in,left=1.25in,top=0.5in,right=1.25in,bottom=1in,includeheadfoot]{geometry}

\usepackage[english]{babel}
\usepackage{tikz}
\usetikzlibrary{positioning,shapes,shadows,arrows}

\usepackage{amsfonts}
\usepackage{amsthm}
\usepackage{amsmath}

\usepackage{booktabs}
\usepackage{mathtools}
\usepackage{amssymb,epsfig,multirow}
\usepackage{amsbsy}
\usepackage{graphicx}
\usepackage{geometry}
\usepackage{url}
\usepackage{float}
\usepackage{color}
\usepackage{enumerate}

\usepackage[pdf]{graphviz}
\usepackage{tikz}
\usetikzlibrary{arrows.meta}
\usepackage{tkz-euclide}
\usepackage{tkz-graph}
\usetikzlibrary{shapes,snakes}
\usetikzlibrary{graphs,quotes,arrows.meta}
\usepackage{subcaption}
\newtheorem{theorem}{Theorem}
\numberwithin{theorem}{section}

\newtheorem{prop}[theorem]{Proposition}

\newtheorem{defn}[theorem]{Definition}%[section]
\newtheorem{rem}[theorem]{Remark}%[section]
\newtheorem{obser}[theorem]{Observation}%[section]

\newcommand{\A}{\mathcal{A}}

\newcommand{\C}{\mathcal{C}}

\newcommand{\PP}{\mathcal{P}}
\renewcommand{\cite}[1]{\parencite{#1}}

\allowdisplaybreaks
\tikzstyle{abstract}=[rectangle, draw=black, rounded corners, fill=blue!40, drop shadow,
text centered, anchor=north, text=white, text width=3cm]
\tikzstyle{comment}=[rectangle, draw=black, rounded corners, fill=red, drop shadow,
text centered, anchor=north, text=white, text width=3cm]
\tikzstyle{myarrow}=[->, >=open triangle 90, thick]
\tikzstyle{line}=[-, thick]

\begin{document}
%\vspace{4.5in}

\title{\LARGE \bf An algorithm with improved complexity for pebble motion/multi-agent path finding on trees}

%\author{S. Ardizzoni$^{1}$, I. Saccani$^{1}$, L. Consolini$^{1}$,  M. Locatelli$^{1}$, B. Nebel$^{2}$  \\
      % { \small $^{1}$Dipartimento di Ingegneria e Architettura, Universit\`a di Parma, Parco Area delle Scienze, 181/A, Parma, Italy}\\
       
    % { \small $^{2}$ Institut für Informatik, Albert-Ludwigs-Universität, Georges-Köhler-Allee 52, Freiburg, Germany
  % }
%}

\author{\name Stefano Ardizzoni \email stefano.ardizzoni@unipr.it \\  
     	\name Irene Saccani \email irene.saccani@unipr.it \\
       \name Luca Consolini \email luca.consolini@unipr.it \\
       \name Marco Locatelli \email marco.locatelli@unipr.it \\
    	\addr Dipartimento di Ingegneria e Architettura, Universit\`a di Parma, \\
       	Parco Area delle Scienze, 181/A, Parma, Italy
    	\AND
    	\name Bernhard Nebel \email nebel@informatik.uni-freiburg.de \\
    	\addr Institut für Informatik, Albert-Ludwigs-Universität \\
	 Georges-Köhler-Allee 52, Freiburg, Germany}

%\date{}

	\maketitle
\begin{abstract}
  The \textit{pebble motion on trees} (PMT) problem consists in finding a feasible sequence of moves that repositions a set of pebbles to assigned target vertices. This problem has been widely studied because, in many cases, the more general Multi-Agent path finding (MAPF) problem on graphs can be reduced to PMT. %One of the most important results about PMT is provided by Kornhauser, who describes a procedure, very difficult to implement, that finds solution of length $O(n^3)$. Other papers present algorithms easier to implement but with worst length complexity.
  We propose a simple and easy to implement procedure, which finds solutions of length $O(knc + n^2)$, where $n$ is the number of nodes, $k$ is the number of pebbles, and $c$ the maximum length of corridors in the tree. This complexity result is more detailed than the current best known result $O(n^3)$, which is equal to our result in the worst case, but does not capture the dependency on $c$ and $k$.
\end{abstract}

\section{Introduction}
\label{Introduction}

Multi-Agent Path-Finding (MAPF), also called pebble motion on graphs, or cooperative path-finding, is the problem of finding a collision-free movement plan for a set of agents (or pebbles) moving on a graph.
This problem has been widely discussed, together with its many variants \cite{MAPF}, on various types of graphs.
For most graph classes, finding an optimal solution of MAPF (that is, a solution with a
minimum number of moves) is NP-hard \cite{yu2013}. Instead, the complexity of checking MAPF feasibility
depends on the specific graph class. For instance,
it is polynomial on undirected graphs \cite{Kornhauser}, on strongly biconnected directed graphs \cite{diBOX}, and on strongly
connected directed graphs \cite{diSC}. Instead, it is NP-hard in the general case of directed graphs \cite{nebel2020}.
Optimal and suboptimal algorithms have been proposed in the last forty years \cite{acha2022,alotaibi2016,diSC,auletta,botea2018,diBOX,wilde2014,Kornhauser,cbs}. 

We focus on one of the simplest versions of this
problem, the \textit{pebble motion on a tree} (PMT) 
\cite{auletta,auletta2001optimal,tree,Tass,PMT}, which is defined as
follows. Let $T=(V,E)$ be a tree with $n$ vertices and $k
< n$ distinct pebbles, numbered $1,\ldots,k$, placed on distinct vertices. A \textit{move} consists in transferring a pebble from its current position to an adjacent unoccupied vertex. The PMT consists in finding a sequence of moves that repositions all pebbles to assigned target vertices. In particular, we focus on the problem of finding a feasible solution, not necessarily optimal.%, but we aim to minimize the length complexity.

Although PMT is one of the simplest versions of MAPF, it is quite relevant. Indeed, various algorithms that solve MAPF on more general graphs are based on a reformulation of MAPF as a PMT, over a suitably defined tree \cite{tree,PMT,diSC}.
We will further discuss this in Section~\ref{tspmt}. In literature, there exist many complete sub-optimal
algorithms for solving PMT. In particular, Kornhauser and coauthors \cite{Kornhauser} present a procedure which solves it in $O(n^3)$ moves. However, the approach is not described algorithmically, but must be derived from a number of proofs in the paper \cite{Roger}. This requires significant effort, and, to the best of our knowledge, Kornhauser's results have never been fully implemented. Auletta and coauthors \cite{auletta} present an algorithm for deciding the feasibility of PMT, from which a solution can be derived requiring $O(k^2(n-k))$ moves. However, the paper does not explicitly explain how to find such a solution. Korshid and coauthors \cite{Tass} present an algorithm for PMT (called TASS) that is easy to understand and implement. However, solutions provided by TASS require $O(n^4)$ moves. 

Therefore, after Kornhauser and coauthors in 1984, no one has proposed a clear and detailed algorithm with length-complexity $O(n^3)$.  The result of Kornhauser is a fundamental step in the study of PMT complexity. However, under a more practical point of view, the implementation of a simple and efficient algorithm for solving PMT with length-complexity at least $O(n^3)$ remains an open problem.  The aim of this paper is to address this problem by proposing an efficient, clear, and simple PMT solver, with a more detailed complexity result with respect to Kornhauser's.

In this work, we also deal with two variants of PMT: the \textit{motion planning} problem and the \textit{unlabeled} PMT. In the first one, a single marked pebble has to reach a desired target vertex, while non-marked pebbles are obstacles
that need be moved out of the way to re-position the
marked one \cite{GMP1R,WuGru09,feasibility,auletta2001optimal}.
In the second one (also known as U-GRAPH-RP  \cite{cualinescu2008,
  dumitrescu2013} or \textit{anonymous MAPF} \cite{anonMAPF}), pebbles are not labeled (i.e., there is a set of target positions $D$, and each pebble has to reach a vertex in $D$, not specified in advance).

\textbf{Statement of contribution.} We present three main contributions:
\begin{enumerate}
	\item In Section \ref{mpt}, we propose the sub-optimal CATERPILLAR algorithm to solve the motion planning problem on a tree. It provides solutions with $O(nc)$ moves, where $c$ is the maximum length of the corridors (i.e., paths whose internal nodes all have degree two, and whose end nodes have degree different from two). We are able to guarantee this complexity since, when we move the marked pebble $p$ to its target, we only move the obstacles that are along the path of $p$, sliding them section by section from one subset of vertices to another, avoiding unnecessary motions.
	
	\item In Section \ref{sec:pmt}, we propose a sub-optimal algorithm for PMT (called \textit{Leaves procedure for} PMT). The idea of the \textit{Leaves procedure for} PMT is to use the CATERPILLAR algorithm to move the pebbles to the leaves of the tree, used as intermediate targets. At the end, we solve an unlabeled PMT instance, which brings the pebbles to the original targets.
	This procedure is simpler and easier to implement than the one proposed by Kornhauser and coauthors \cite{Kornhauser}.  In addition, we prove a more detailed complexity result than the one provided by Kornhauser and coauthors. Indeed, our algorithm finds solutions
        with a number of moves $O(k n c+n^2)$, which in the worst case
        is $O(n^3)$. Therefore, the number of moves of these solutions
        depends on the tree structure and the number of pebbles.  %In addition, they have a better quality in terms of length than those 

       % Furthermore, the solutions provided by this algorithm have the interesting property that each vertex is crossed at most $O(kc)$ times by the pebbles.

	\item In Section \ref{tspmt}, we discuss a variant of the PMT problem, called PMT \textit{with trans-shipment vertices} (\textit{ts}-PMT). This variant is relevant since a MAPF instance on a generic graph can be
        reduced to an instance of this problem. \textit{ts}-PMT can be solved with the \textit{Leaves
          procedure for} PMT with some minor
        modifications. This permits us to provide an upper
          bound for the solution length of MAPF on graphs.
\end{enumerate}

\section{Problem Definition}

%To define the problem we deal with in this paper, we need some important concepts.
Let $T=(V,E)$ be a tree, with vertex set $V$ and edge set $E$. We are also given a set $P$ of pebbles and a set $H$ of holes, and
each vertex of $T$ is occupied  either by a pebble or by a hole, so that $|V| = |P|+|H|$.
A \textit{configuration} is a function  $\A:P \cup H \rightarrow V$ that assigns to each pebble or hole the vertex occupied by it. A configuration is \textit{valid} if it is one-to-one (i.e., each vertex is occupied by one and only one pebble or hole). The collection $\C \subset \{P \cup H \to V\}$ contains all valid configurations. %Set $H(G,\A)=\A(H)$.
%Since holes represent unoccupied vertices, we are 
%we are more interested about pebbles positions.

%We represent the restiction of a configuration $\A$ to the pebbles as
%\[ \pi(\A):= \A_{|P}.
%\]

%Let $W \subset V$ be a subset of $V$, the \textit{inverse image of $W$ with repsect to $\A$} is $\A^{-1}(W)= \{x \in P \cup H: \A(x) \in W\}$.

%\A_{|\A^{-1}(W)}$.  

Given a configuration $\A$ and $u, v \in V$, we denote by $\A[u,v]$ the configuration obtained from $\A$ by exchanging the pebbles (or holes) placed at $u$ and $v$:
%\vspace{-2pt}
\begin{equation} 
	\label{c}
	\A[u,v](q):=  \Bigg\{
	\begin{array}{ll}       	
		
		v, & \text{if } \A(q)  = u ;\\
		
		u, & \text{if } \A(q)  = v ;\\
		
		\A(q), & \text{otherwise } .\\
	\end{array}
\end{equation}
%Moreover, given a pebble $q \not \in P$ and $u \in H(G,\A)$ we can define an \textit{extension $\A[q^u]$ of} $\A$ as follows:
% \begin{equation} 
	%\label{ext}
	%\bar \A[q^u](p):=  \Big\{
	%\begin{array}{lr}
	%\A(p), & \text{if } p \in P;\\
	
	%u, & \text{if } p  = q.\\
	%\end{array}                                                                                                                                                                                                                                                                                                                                                                                               
	%\end{equation}

	As mentioned in the Introduction, a {\em move} is the movement of a pebble along an edge.  For each edge $(u,v) \in E$ we can define two possible moves, that are the two ordered pairs $u \rightarrow v$ and $v \rightarrow u$. We indicate with $\hat{E}$ the set of all the moves on tree $T$.
	Function $\rho: \C \times \hat{E} \rightarrow \C$ is a partially defined transition function such that $\rho(\A,u \rightarrow v)$ is defined if and only if $v$ is empty (i.e., occupied by a hole). In this case $\rho(\A,u \rightarrow v)$ is the configuration obtained by exchanging the pebble or the hole in $u$ with the hole in $v$. Notation $\rho(\A,u \rightarrow v)!$ means that the function is defined. In other words
	$\rho(\A,u \rightarrow v)!$ if and only if $(u,v)\in E$ and $\A^{-1}(v) \in H$. If $\rho(\A, u \rightarrow v)!$, then
	$\rho(\A, u \rightarrow v)=\A[u,v]$. Note that the hole in $v$ moves along $v \rightarrow u$, while the pebble or hole on $u$ moves on $v$.
	
	%\[ \rho(\A, u \rightarrow v):=  \Big\{
	%        	\begin{array}{ll}       	       		
		%        		\A, & \text{if }  v\in \A(P);\\
		%        			\A[u,v], & \text{otherwise } .\\
		%        	\end{array}
	%       \]
	
	We represent a \textit{plan} as an ordered sequence of moves.
	It is convenient to view the elements of $\hat{E}$ as the symbols of a language.
	We denote by $E^*$ the Kleene star of $\hat{E}$, that is the set of
	ordered sequences of elements of $\hat{E}$ with arbitrary length, together with the empty string $\epsilon$:
	\[
E^*=\bigcup_{i=1}^\infty \hat{E}^i \cup \{\epsilon\}.
	\]

	We extend the function $\rho:\C \times \hat{E} \to \C$ to $\rho: \C \times E^*  \to \C$, by setting $(\forall \A \in \C) \rho(A,\epsilon)!$ and $\rho(\A,\epsilon)=\A$. Moreover,  $(\forall s \in E^*, e \in \hat{E}, \A \in \C)$  $\rho(\A,  se) !$ if and only if $\rho(\A, s)!$ and $\rho(\rho(\A, s), e)!$ and, if $\rho(\A se)!$, then $\rho(\A se)=\rho(\rho(\A s),e)$.
	Note that $\epsilon$ is the trivial plan that keeps all pebbles and holes on their positions. We denote by $|f|$ the length of a plan $f$ (i.e., the number of moves of $f$).
	We define an equivalence relation $\sim$ on $E^*$, by setting, for $s,t \in E^*$, $s \sim t \leftrightarrow (\forall \A \in \C)\, \rho(\A,s)=\rho(\A,t)$.
	In other words, two plans are equivalent if they reconfigure pebbles and holes in the same way. Given a configuration $\A$ and a plan $f$ such that $\rho(\A, f)!$, a plan $f^{-1}$ is a \emph{reverse} of $f$ if $\A =\rho(\rho(\A, f),f^{-1})$  (i.e., $f^{-1}$ moves each pebble and hole back to their initial positions). We can also write $f f^{-1}\sim \epsilon$, so that $f^{-1}$ behaves like a right-inverse.\\

	\begin{prop}
		\label{reverse}
                For any configuration $\A \in \C$ and any plan $f \in E^*$, such that $\rho(\A,f)!$, there exists a reverse plan $f^{-1}$ such that $|f| = |f^{-1}|$.
\end{prop}
\begin{proof}
	The thesis can be proved by induction as follows.
If $f \in \hat{E}$, then there exist $u,v \in V$ such that $f = u \rightarrow v$ and $f^{-1} = v \rightarrow u$. Now, suppose that for any $f \in E^*$ such that $|f|= n >1$ there exists a reverse $f^{-1}$ such that $|f|=|f^{-1}|$. We prove that 
the thesis is verified also for each plan of length $n+1$. Indeed, if $|f| = n+1$ then $f= ge$ with $|g| = n$ and $e \in \hat{E}$. Therefore, there exist $g^{-1}$ and $e^{-1}$ the corresponding reverse plans such that $|g^{-1}|=|g|$ and $|e^{-1}|=|e|$, and we can define $f^{-1}:=e^{-1}g^{-1}$ which is a reverse plan of $f$:		
\[  ff^{-1} = gee^{-1}g^{-1} \sim g g^{-1} \sim \epsilon.\]
Moreover, $|f^{-1}|=|g^{-1}|+|e^{-1}| = n + 1$.
\end{proof}

	\vspace{5mm}
	
	%	 Moreover, given $u,v \in V$, we define \textit{exchange plan} as the plan $f_{uv}$, if exists, such that $\A[u,v]= \rho(\A,f_{uv})$.
	
	%An agent can be moved in discrete time steps.
	%(DOVE METTERE ???)For simplicity, in this theoretical study we assume
	%that agents move one at a time.

Our main problem is \textit{pebble motion on a tree}, which consists in finding a plan that re-positions all pebbles to assigned target vertices, avoiding collisions. 
	For this problem, and the ones we will present later, the position of the holes is not relevant. Therefore, we introduce an equivalence relation $\dot{\sim}$ between configurations 
	\[\A^1 \dot{\sim} \A^2 \iff \forall \, p \in P \quad  \A^1(p) = \A^2(p),\]
	and we indicate with $\tilde{\A}^1$ the equivalence class of $\A^1$.

		\begin{defn}{(\textbf{PMT problem}).}
		Given a tree $T$, a pebble set $P$, an initial valid configuration $\tilde{\A}^s$, and a final valid configuration $\tilde{\A}^t$, find a plan $f$ such that $\tilde{\A}^t= \rho(\tilde{\A}^s,f)$.
	\end{defn}
	We also focus on two relaxations of the PMT problem: the \textit{ motion planning problem } and the \textit{ unlabeled PMT problem}. The former consists in finding a plan such that a
	single marked pebble reaches a desired target vertex, avoiding
	collisions with other pebbles, which are movable \textit{obstacles}. The latter consists  in finding a plan such that each pebble reaches any vertex belonging to the set of targets.

		\begin{defn}{(\textbf{Motion planning problem}).}
		Let $T=(V,E) $ be a tree, $P$ a set of pebbles, and $\tilde{\A}^s$ an initial valid configuration. Given a pebble $\bar{p}$ and a target node $t\in V$, find a plan $f$ such that $t = \rho(\tilde{\A}^s,f)(\bar{p})$.
	\end{defn}

	\begin{defn}{(\textbf{Unlabeled PMT problem}).}
		Let $T=(V,E) $ be a tree, $P$ a set of pebbles, and $\tilde{\A}^s$ an initial valid configuration.
Given $D \subset V$, a set of destinations such that $|D|=|P|$, find a plan $f$ such that $D = \rho(\tilde{\A}^s,f)(P)$.
\end{defn}

\subsection{Notation}
Let $T=(V,E)$ be a tree with $n$ nodes and let $u\in V$. We denote by $F(u)$ the connected components of the \textit{forest} obtained from $T$ by deleting $u$. For some $v\in V\setminus\{u\}$, $T(F(u),v)$ is the connected component containing $v$, while $C(F(u), v)$ is the set of remaining connected components of $F(u) $ excluding $T(F(u),v)$.

Given two nodes $a,b\in V$, we denote by $\pi_{ab}$ the set of vertices of the unique path in $T$ from $a$ to $b$, and with $d(a,b)$ the length of this path. 
In particular, $\pi_{ab}$ is a \textit{corridor} if $a$ and $b$ have degree different from two, while the internal nodes of the path all have exactly degree two. Moreover, $\pi_{ab} \setminus \{a\}$ is the set of all vertices of $\pi_{ab}$, except for $a$. We denote by $C(T)$ the set of all corridors of $T$, and by $c_1$ the maximum corridor length:

\[  c_1 (T)= \max \{d(a,b) : \, \pi_{ab} \in C(T)\}.\]

Let $J\subset V$ be the set of \textit{junctions} (i.e., nodes with degree greater than two). We define
the subclass of corridors $\bar{C}(T) \subset C(T)$  that have only junctions as endpoints, and we denote by $c_2(T)$ the maximum length of this subclass of corridors. Obviously, $c_2(T) \leq c_1(T)$.  Note that corridors in $C(T) \setminus \bar{C}(T)$ are those for which at least one endpoint is a leaf of the tree.\\

Let us define the distance between a node $s$ and a subset of nodes $W \subset V$ as 
\[ d(s,W) = \min_{a \in W} \; d(s,a).\]
Furthermore, we define the distance of $W_1$ from $W_2$  as 
\[ d(W_1,W_2) = \sum_{a \in W_1}   d(a,W_2)   .  \]

\noindent  Then, a \textit{subset of $V$ of cardinality $q$ closest  to $W_1$} is a set $W$ such that

\[
W \in \arg \min_{\begin{array}{c} \scriptstyle W \in \PP(V) \\ 	
		\scriptstyle 	|W| = |q|\end{array}
} d(W,W_1),
\]
where $\PP(V)$ denotes the power set of $V$ (i.e., the set of all its subsets).

\subsection{Assumptions}
\label{sec:assum}

In all the problems we focus on, we assume that the following assumption holds:

%\begin{enumerate}
%	\item $J \not = \emptyset$, i.e., there is at least one junction;
\begin{equation}\label{ass}
	|H| \geq c(T),
	\end{equation}
%\end{enumerate}
where $c(T) \in \mathbb{N}$ is a constant depending on the structure of the tree $T$. In particular, if $T$ is a path graph (i.e., a tree with two nodes of degree 1, and the remaining $n-2$ nodes of degree 2) $c(T) := c_1(T)$. 
Otherwise, in all other cases, $c(T)$ is defined as follows:
\begin{equation}\label{c}
	c(T):=\max \{c_1(T)+1,\, c_2(T) +2\}.
\end{equation}

\noindent From now on, when we write $c_1$, $c_2$ and $c$ without the indication of a tree within the parenthesis, it is given as understood that the three parameters refer to the tree $T$ on which we are solving the PMT problem, while for other trees we will explicitly indicate them within parenthesis .

Kornhauser and coauthors \cite{Kornhauser} showed that Assumption (\ref{ass}) is a necessary and sufficient condition for the feasibility of PMT on trees. 
Obviously, it follows that this condition is also sufficient for the feasibility of any instance of the unlabeled PMT problem, and of the motion planning problem.

\subsection{Basic plans}
\begin{figure}[h!]
	\centering
%	\resizebox{\columnwidth}{!}{
	 \begin{subfigure}[t]{0.5\textwidth}
	 	\centering
		\begin{tikzpicture}
			[scale=1.2,auto=left]
			\node[style={circle,fill=blue!30}] (A) at (0,0) {\textbullet};
			\node[style={rectangle,fill=green!40}] (B) at (1.5,0) {\textbullet};
			\node[below] at (0,-0.2) {$w \equiv u_5$};
			\node[style={rectangle,fill=green!40}] (J) at (1.5,1.5) {\textbullet};
			\node[below] at (1.5,-0.2) {$u_4$};
			\node[style={rectangle,fill=green!40}] (C) at (3,0) {\textbullet};
				\node[below] at (3.2,-0.2) {$u_3$};
			%	\node[below] at (A) {$s$};
			%	\node[above] at (A) {$f_0$};
			%	\node[below] at (B) {$i_0$};
			%	\node[above] at (C) {$i_1$};
			\node[style={circle,fill=blue!30}] (K) at (3,-1.5) {\textbullet};
			
			\draw  (1.5,0) -- (1.5,1.5);
			\draw  (3,0) -- (3,-1.5);
			%\node[left] at (K) {$f_1$};
			\node[style={rectangle,fill=green!40}] (D) at (4.5,0) {\textbullet};
			%	\node[above] at (D) {$z_2$};
			\node[style={rectangle,fill=green!40}] (E) at (6,0) {\textbullet};
			%	\node[below] at (6.15,0) {$s$};
		\node[below] at (4.5,-0.2) {$u_2$};
			\node[below] at (6,-0.2) {$v \equiv u_1$};

			\draw  (0,0.05) -- (6,0.05);
		
				\path[->]  (B)  edge[bend right=40,draw=red] node[above]{{\color{red}$(1)$}}  (A);
		\path[->]  (C)  edge[bend right=40,draw=red]   node[above]{{\color{red}$(2)$}} (B);
			\path[->]  (D)  edge[bend right=40,draw=red]  node[above]{{\color{red}$(3)$}}  (C);
			\path[->]  (E)  edge[bend right=40,draw=red]  node[above]{{\color{red}$(4)$}}  (D);
		\end{tikzpicture}
	\caption{Initial configuration. The red edges represent plan $\alpha_{vw}=(u_{4}\rightarrow w,u_{3}\rightarrow u_4,u_{2}\rightarrow u_3,v\rightarrow u_{2})$.}
	
	\end{subfigure}

\vspace{5mm}

\begin{subfigure}[t]{0.5\textwidth}
	\centering
			\begin{tikzpicture}
		[scale=1.2,auto=left]
		\node[style={rectangle,fill=green!40}] (A) at (0,0) {\textbullet};
		\node[style={rectangle,fill=green!40}] (B) at (1.5,0) {\textbullet};
		\node[below] at (0,-0.2) {$w\equiv u_5$};
		\node[style={rectangle,fill=green!40}] (J) at (1.5,1.5) {\textbullet};
		\node[below] at (1.5,-0.2) {$u_4$};
		\node[style={rectangle,fill=green!40}] (C) at (3,0) {\textbullet};
		\node[below] at (3.2,-0.2) {$u_3$};
		%	\node[below] at (A) {$s$};
		%	\node[above] at (A) {$f_0$};
		%	\node[below] at (B) {$i_0$};
		%	\node[above] at (C) {$i_1$};
		\node[style={circle,fill=blue!30}] (K) at (3,-1.5) {\textbullet};
		
		\draw  (1.5,0) -- (1.5,1.5);
		\draw  (3,0) -- (3,-1.5);
		%\node[left] at (K) {$f_1$};
		\node[style={rectangle,fill=green!40}] (D) at (4.5,0) {\textbullet};
		%	\node[above] at (D) {$z_2$};
		\node[style={circle,fill=blue!30}] (E) at (6,0) {\textbullet};
		%	\node[below] at (6.15,0) {$s$};
		\node[below] at (4.5,-0.2) {$u_2$};
		\node[below] at (6,-0.2) {$v \equiv u_1$};

		\draw  (0,0.05) -- (6,0.05);

	\end{tikzpicture}
%	}
\caption{Final configuration after bringing the hole from $w$ to $v$.}

\end{subfigure}
\vspace{5mm}

	\caption{Example of \textsc{Bring hole from $w$ to $v$}. Green squares represent pebbles, blue circles represent holes.}
	\label{bring}

\end{figure}
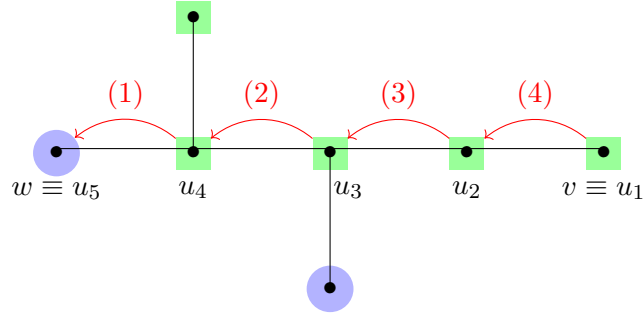
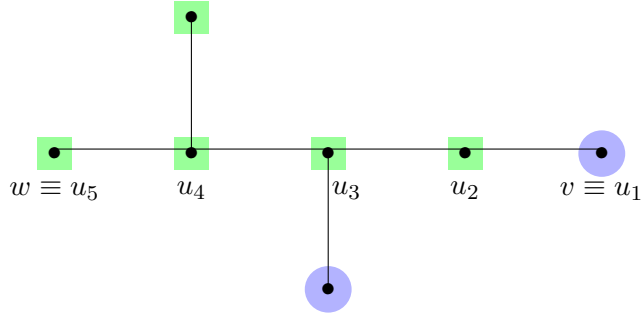

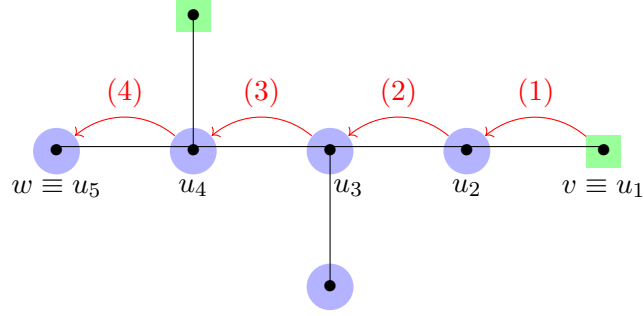
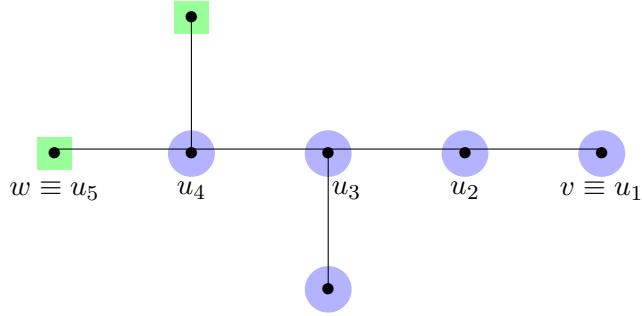
\begin{figure}[h!]
	\centering
	\begin{subfigure}[t]{0.5\textwidth}
		\centering
	%	\resizebox{\columnwidth}{!}{
		\begin{tikzpicture}
			[scale=1.2,auto=left]
			\node[style={circle,fill=blue!30}] (A) at (0,0) {\textbullet};
			\node[style={circle,fill=blue!30}] (B) at (1.5,0) {\textbullet};
			\node[below] at (0,-0.2) {$w\equiv u_5$};
			\node[style={rectangle,fill=green!40}] (J) at (1.5,1.5) {\textbullet};
			\node[below] at (1.5,-0.2) {$u_4$};
			\node[style={circle,fill=blue!30}] (C) at (3,0) {\textbullet};
			\node[below] at (3.2,-0.2) {$u_3$};
			%	\node[below] at (A) {$s$};
			%	\node[above] at (A) {$f_0$};
			%	\node[below] at (B) {$i_0$};
			%	\node[above] at (C) {$i_1$};
			\node[style={circle,fill=blue!30}] (K) at (3,-1.5) {\textbullet};
			
			\draw  (1.5,0) -- (1.5,1.5);
			\draw  (3,0) -- (3,-1.5);
			%\node[left] at (K) {$f_1$};
			\node[style={circle,fill=blue!30}] (D) at (4.5,0) {\textbullet};
			%	\node[above] at (D) {$z_2$};
			\node[style={rectangle,fill=green!40}] (E) at (6,0) {\textbullet};
			%	\node[below] at (6.15,0) {$s$};
			\node[below] at (4.5,-0.2) {$u_2$};
			\node[below] at (6,-0.2) {$v \equiv u_1$};

			\draw  (0,0.05) -- (6,0.05);
			
			\path[->]  (B)  edge[bend right=40,draw=red] node[above]{{\color{red}$(4)$}}  (A);
			\path[->]  (C)  edge[bend right=40,draw=red]   node[above]{{\color{red}$(3)$}} (B);
			\path[->]  (D)  edge[bend right=40,draw=red]  node[above]{{\color{red}$(2)$}}  (C);
			\path[->]  (E)  edge[bend right=40,draw=red]  node[above]{{\color{red}$(1)$}}  (D);
		\end{tikzpicture}
	
		\caption{Initial configuration. The red edges represent plan 	$\beta_{vw} = (v\rightarrow u_{2}, u_2\rightarrow u_{3},u_3\rightarrow u_{4}, u_{4}\rightarrow w)$.}
	
\end{subfigure}

\vspace{5mm}

\begin{subfigure}[t]{0.5\textwidth}
\centering
		\begin{tikzpicture}
			[scale=1.2,auto=left]
			\node[style={rectangle,fill=green!40}] (A) at (0,0) {\textbullet};
			\node[style={circle,fill=blue!30}] (B) at (1.5,0) {\textbullet};
			\node[below] at (0,-0.2) {$w\equiv u_5$};
			\node[style={rectangle,fill=green!40}] (J) at (1.5,1.5) {\textbullet};
			\node[below] at (1.5,-0.2) {$u_4$};
			\node[style={circle,fill=blue!30}] (C) at (3,0) {\textbullet};
			\node[below] at (3.2,-0.2) {$u_3$};
			%	\node[below] at (A) {$s$};
			%	\node[above] at (A) {$f_0$};
			%	\node[below] at (B) {$i_0$};
			%	\node[above] at (C) {$i_1$};
			\node[style={circle,fill=blue!30}] (K) at (3,-1.5) {\textbullet};
			
			\draw  (1.5,0) -- (1.5,1.5);
			\draw  (3,0) -- (3,-1.5);
			%\node[left] at (K) {$f_1$};
			\node[style={circle,fill=blue!30}] (D) at (4.5,0) {\textbullet};
			%	\node[above] at (D) {$z_2$};
			\node[style={circle,fill=blue!30}] (E) at (6,0) {\textbullet};
			%	\node[below] at (6.15,0) {$s$};
			\node[below] at (4.5,-0.2) {$u_2$};
			\node[below] at (6,-0.2) {$v \equiv u_1$};

			\draw  (0,0.05) -- (6,0.05);

		\end{tikzpicture}
		%	}
	
	\caption{Final configuration after moving the pebble from $v$ to $w$. }
	
\end{subfigure}
\vspace{5mm}

	\caption{Example of \textsc{Move Pebble from $v$ to $w$}. Green squares represent pebbles, blue circles represent holes.}
	\label{move}
\end{figure}

Let $T$ be a tree and $\A$ a configuration on it. Given a path $\pi_{vw}=v \, u_2\, \cdots \, u_{n-1} \, u_n\equiv w$, we define the following plans:
\begin{enumerate}
	\item If $w \in \A(H)$, \textsc{Bring hole from $w$ to $v$} is defined as
	\begin{equation}
		\label{eq:bringhole1}
		\alpha_{vw} = (u_{n-1}\rightarrow w,\ldots,v\rightarrow u_{2}).
	\end{equation}
	In other words, for each $j$ from $n-1$ to $1$, if there is a pebble on $u_j$, we move it on $u_{j+1}$. For instance, see the example of Figure \ref{bring}.
	\item If $v \in \A(P)$ and $\pi_{vw} \setminus \{v\} \subset \A(H)$, \textsc{Move Pebble from $v$ to $w$} is defined as
	\begin{equation}
		\label{eq:bringhole2}
		\beta_{vw} = (v\rightarrow u_{2},  \ldots, u_{n-1}\rightarrow w).
	\end{equation}
 For instance, see the example of Figure \ref{move}.
\end{enumerate}

	\section{Unlabeled PMT problem}
\label{sec:upmt}
	Let $P$ be a set of unlabeled pebbles on a tree $T=(V,E)$.
	Let $\A^s$ be the initial configuration and $D$ be the set of destinations. We denote by $S=\A^s(P)$ the set of pebbles initial positions.
	The goal of the \textit{unlabeled pebble motion on trees} is to move each pebble from its initial position to any position of $D$. In the following, we introduce an algorithm presented by Kornhauser and coauthors \cite{Kornhauser}.

	\begin{enumerate}
		\item If $V$ is empty: terminate the procedure. 
		
		\item Select any leaf $v$ of $T$. 
		\begin{itemize}
			\item If $v\in S\cap D$ or $v \not \in  S \cup D$, then "prune" $v$ from $T$, i.e., $V=V \setminus \{v\}$, and set $S=S\setminus \{v\}$. Go to Step 1.
			
			\item If $v \in D \setminus S$, select a pebble $p$ on a vertex $w$ such that				
			\[
			w \in \arg \min_{v' \in S
			} d(v',v).
			\]
			
			Let $p$ be the pebble on $w$. By definition of $w$, path $\pi_{wv}$ contains only pebble $p$. Therefore, move $p$ to $v$ and 
			update $S= S \setminus \{w\}$. Then, "prune" $v$ from $T$, i.e., $V=V \setminus \{v\}$. Go to Step 1.
			
			\item if $v \in S \setminus D$. Find an unoccupied vertex $u$ which is at minimum distance from $v$ on $T$:

			\[
			u \in \arg \min_{v' \in V \setminus S
			} d(v',v).
			\]
			
			 Then, path $\pi_{vu}$  has pebbles on each vertex except  $u$. Move each pebble on the path $\pi_{vu}$ towards $u$ with plan $\alpha_{vu}$ as defined in (\ref{eq:bringhole1}). This makes $v$ unoccupied and $u$ occupied.
			 Then, set $S= (S \setminus \{v\}) \cup \{u\} $, "prune" $v$ from $T$, i.e., $V=V \setminus \{v\}$, and go to Step 1.
		\end{itemize}
		
	\end{enumerate}
	Since at most $n$ moves are made at each execution of Step 2, and Step 2 is executed $n$ times (at each iteration the cardinality of $V$ is decreased by one), the total number of required moves is at most $n^2$. Therefore, the complexity of this algorithm is $O(n^2)$. \\

	\subsection{Gather holes problem}
	\label{sec_gather}
	In this subsection, we focus on a particular case of the \textit{unlabeled} PMT problem: the \textit{gather holes problem}.
	Let $T=(V,E)$ be a tree with $n$ nodes, $P$ a set of pebbles, and $H$ the set of holes. 
Let $\bar{T}=(\bar{V},\bar{E})$ be a subtree with $q= |\bar{V}|\leq |H|$. Then,  \textit{gather holes in $\bar{T}$} consists in bringing $q$ holes of the tree to the nodes of $\bar{T}$. 
	\begin{defn}{(\textbf{Gather holes problem}).}
	Let $T $ be a tree and $\bar{T}=(\bar{V},\bar{E})$ be a subtree. Let $P$ be a set of pebbles, and $\tilde{\A}^s$ an initial valid configuration.
Find a plan $f$ such that $\bar{V} \cap \rho(\tilde{\A}^s,f)(P) = \emptyset$.
\end{defn}

Even if \emph{gather-holes} can be solved by the previous algorithm, we present a specific procedure that allows finding a feasible solution with a lower time-complexity.
The solution plan removes pebbles from vertices of $\bar{T}$, and replaces them with holes. To search for a short plan, it is convenient to bring holes that are already close to $\bar{V}$. Therefore, we choose the holes in a set $M$ such that
\begin{equation}
\label{eq:choiceM}
M\in \arg\min_{W\subset \A^s(H):\ |W|=q} d(W,\bar{V}),
\end{equation}
i.e., $M$ is a subset of vertices with cardinality $q$ closest to $\bar{V}$ and containing only holes of the initial configuration. Denote by $\tilde{H}$ the set of holes on $M$ ($\A^s(\tilde{H})=M$).
Then, we want to find a plan $f$ such that 
	$\bar{V} = \rho(\A^s,f)(\tilde{H})$.
Moreover, let $\bar H=\{h \in \tilde H: \A^s(h) \notin \bar V\}$.	
	
	 Denoting by $ \bar{V}_{P} = \A^s(P) \cap \bar{V}$ the initial set of occupied vertices of $\bar{T}$,  we can proceed as follows:
	
	\begin{enumerate}
		\item If $\bar{V}_{P}$  is empty: terminate the procedure;
		\item Select $h \in \bar{H}$,  let  $v =\A^s(h)$  and  $u \in \bar{V}_{P}$ be a closest node of $\bar{V}_{P}$ to $v$:
		
		\begin{equation}
\label{eq:choiceu}
		u \in \arg \min_{v' \in \bar{V}_{P}
		} d(v',v).
		\end{equation}
		
		 Denote by $p$ the pebble on $u$. If $\pi_{uv} \cap \bar{V} = \{u\}$, then move each pebble on the path $\pi_{uv}$ towards $v$ with plan $\alpha_{uv}$ defined in (\ref{eq:bringhole1}), and update $\A^s = \rho(\A^s, \alpha_{uv})$. Otherwise, let $w$ be the closest node to $v$ of $\pi_{uv} \cap \bar{V} $:  move $p$ from $u$ to $w$ with plan $\beta_{uw}$ defined in (\ref{eq:bringhole2}) (note that $\pi_{uw} \setminus \{u\}$ contains only unoccupied vertices); then, move each pebble on path $\pi_{wv}$ towards $v$ with plan $\alpha_{wv}$. Finally, update $\A^s = \rho(\A^s, \beta_{uw} \alpha_{wv})$. This makes $u$ unoccupied.
		
		\item Update $\bar{H}= \bar{H} \setminus \{h\}$ and $\bar{V}_{P}= \bar{V}_{P} \setminus \{u\}$. Go to Step 1.
	\end{enumerate}

		Since  at most $n$ moves (with $n=|V|$) are
                  made at each execution of Step 2, and Step 2 is
                  executed at most $q$ times (with $q=|\bar{V}|$), (since the cardinality of $\bar{V}_P$ is reduced by one at each iteration), we have the following complexity result.
\begin{prop}
\label{prop:gatherhole}
The length complexity of the solution provided by the gather holes procedure is $O(n q)$.
\end{prop} 
		
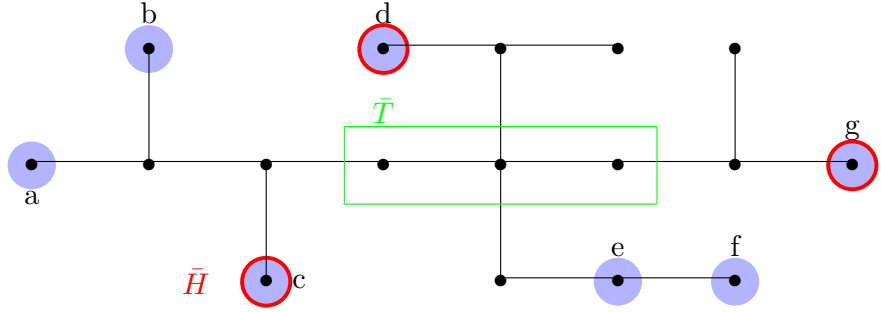
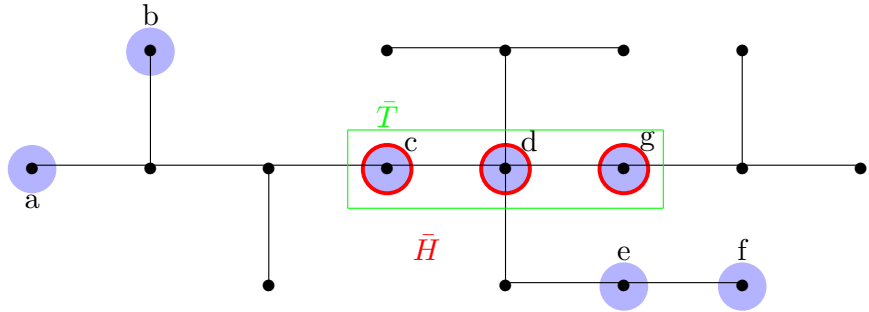
\begin{figure}[h!]
				\centering
			\begin{subfigure}[t]{0.8\columnwidth}
				\centering
			\resizebox{\columnwidth}{!}{
				\begin{tikzpicture}
				%	[scale=0.8]
					\node[style={circle,fill=blue!30}] (A) at (0,0) {\textbullet};
					\node (B) at (1.5,0) {\textbullet};
						\node[below] at (0,-0.2) {a};
					\node[style={circle,fill=blue!30}] (J) at (1.5,1.5) {\textbullet};
						\node[above] at (1.5,1.7) {b};
					\node (C) at (3,0) {\textbullet};
					%	\node[below] at (A) {$s$};
					%	\node[above] at (A) {$f_0$};
					%	\node[below] at (B) {$i_0$};
					%	\node[above] at (C) {$i_1$};
					\node[style={circle,fill=blue!30},shape=circle,draw=red,line width=0.5mm] (K) at (3,-1.5) {\textbullet};
						\node[right] at (3.2,-1.5) {c};
					\draw  (1.5,0) -- (1.5,1.5);
					\draw  (3,0) -- (3,-1.5);
					%\node[left] at (K) {$f_1$};
					\node(D) at (4.5,0) {\textbullet};
				%	\node[above] at (D) {$z_2$};
					\node (E) at (6,0) {\textbullet};
				%	\node[below] at (6.15,0) {$s$};
					\node (F) at (7.5,0) {\textbullet} ;
					
					\node[style={circle,fill=blue!30}]   (L) at (7.5,-1.5) {\textbullet};
					\node[above] at (7.5,-1.3) {e};
					%	\node[left] at (L) {$f_2$};
					%\draw (7.5,0) -- (7.5,-1.5);
					%	\node[above] at (F) {$j_1=i_2$};
					\node (G) at (9,0) {\textbullet} ;
					\draw  (9,0) -- (9,1.5);
					\node (M) at (9,1.5) {\textbullet};
					\node[style={circle,fill=blue!30},shape=circle,draw=red,line width=0.5mm]  (H) at (10.5,0) {\textbullet} ;
						\node[above] at (10.5,0.2) {g};
					%\node (I) at (12,0) {\textbullet} ;
					%\node[above] at (I) {$t=j_2$};
					%	\draw  (12,0) -- (12,-1.5);
					%	\node (N) at (12,-1.5) {\textbullet};
					\node (I) at (6,1.5) {\textbullet};
				%	\node[above] at (I) {$z_3$};
					\node[style={circle,fill=blue!30},shape=circle,draw=red,line width=0.5mm]  (N) at (4.5,1.5) {\textbullet};
					
						\node[above] at (4.5,1.7) {d};
					\node (O) at (7.5,1.5) {\textbullet};
					\draw  (0,0.05) -- (10.5,0.05);
					\draw  (4.5,1.55) -- (7.5,1.55);
					\draw  (6,0) -- (6,1.5);
					
					\node (P) at (6,-1.5) {\textbullet};
					\node[style={circle,fill=blue!30}] (Q) at (9,-1.5) {\textbullet};
					\node[above] at (9,-1.3) {f};
					\draw (6,-1.45) -- (9,-1.45);
					\draw (6,0) -- (6,-1.5);
				%	\node[below] at (P) {$z_1$};
				%	\node[above] at (10.5,0.2) {$t$};
					%\path [->] (A)  edge[bend right=60]  node[above] {$0$} (D);
					%\path [->] (A)  edge[bend right=80]  node[above] {$0$} (E);
					\draw[draw = green] (4,0.5) -- (8,0.5);
					
					\draw[draw = green] (4,-0.5) -- (8,-0.5);
					\draw[draw = green] (4,0.5) -- (4,-0.5);
					\draw[draw = green] (8,0.5) -- (8,-0.5);
					\node (O) at (4.5,0.7) {{\color{green} $\bar{T}$}};
						\node (O) at (2.1,-1.5) {{\color{red} $\bar{H}$}};
					%\path [->] (B)  edge[bend left=80]  node[above] {$\beta_1$} (E);

					%	\path  (F) edge  (I);
					%	\path [->] (E)  edge[bend left=40]   (H);
					%	\path [->] (E)  edge[bend right=40]   (G);
					%	\path [->] (E) edge  (F);
					%	\path [->] (G)  edge[bend right=40]   (I);
					%	\path [->] (H)  edge[bend left=40]   (I);

				\end{tikzpicture}
			}
			\caption{Initial configuration. We choose $\bar{H}=\{c,d,g\}$ the subset of holes closest to subtree $\bar{T}$.}
			\label{gather1}

\end{subfigure}

\vspace{5mm}

\begin{subfigure}[t]{0.8\columnwidth}
\centering

	\resizebox{\columnwidth}{!}{
		\begin{tikzpicture}
			%	[scale=0.8]
			\node[style={circle,fill=blue!30}] (A) at (0,0) {\textbullet};
			\node[below] at (0,-0.2) {a};
			\node (B) at (1.5,0) {\textbullet};
				\node[above] at (1.5,1.7) {b};
			\node[style={circle,fill=blue!30}] (J) at (1.5,1.5) {\textbullet};
			\node (C) at (3,0) {\textbullet};
			%	\node[below] at (A) {$s$};
			%	\node[above] at (A) {$f_0$};
			%	\node[below] at (B) {$i_0$};
			%	\node[above] at (C) {$i_1$};
			\node (K) at (3,-1.5) {\textbullet};
			
			\draw  (1.5,0) -- (1.5,1.5);
			\draw  (3,0) -- (3,-1.5);
			%\node[left] at (K) {$f_1$};
			\node[style={circle,fill=blue!30},shape=circle,draw=red,line width=0.5mm] (D) at (4.5,0) {\textbullet};
			%	\node[above] at (D) {$z_2$};
			\node[style={circle,fill=blue!30},shape=circle,draw=red,line width=0.5mm] (E) at (6,0) {\textbullet};
			%	\node[below] at (6.15,0) {$s$};
			\node[style={circle,fill=blue!30},shape=circle,draw=red,line width=0.5mm] (F) at (7.5,0) {\textbullet} ;
				\node[above] at (4.8,0.1) {c};
				\node[above] at (6.3,0.1) {d};
				\node[above] at (7.8,0.1) {g};
			\node[style={circle,fill=blue!30}]   (L) at (7.5,-1.5) {\textbullet};
			%	\node[left] at (L) {$f_2$};
			%\draw (7.5,0) -- (7.5,-1.5);
			%	\node[above] at (F) {$j_1=i_2$};
			\node (G) at (9,0) {\textbullet} ;
			\draw  (9,0) -- (9,1.5);
			\node (M) at (9,1.5) {\textbullet};
			\node  (H) at (10.5,0) {\textbullet} ;
			%\node (I) at (12,0) {\textbullet} ;
			%\node[above] at (I) {$t=j_2$};
			%	\draw  (12,0) -- (12,-1.5);
			%	\node (N) at (12,-1.5) {\textbullet};
			\node (I) at (6,1.5) {\textbullet};
			%	\node[above] at (I) {$z_3$};
			\node  (N) at (4.5,1.5) {\textbullet};
			\node (O) at (7.5,1.5) {\textbullet};
			\draw  (0,0.05) -- (10.5,0.05);
			\draw  (4.5,1.55) -- (7.5,1.55);
			\draw  (6,0) -- (6,1.5);
			
			\node (P) at (6,-1.5) {\textbullet};
			\node[style={circle,fill=blue!30}] (Q) at (9,-1.5) {\textbullet};
			\draw (6,-1.45) -- (9,-1.45);
			\draw (6,0) -- (6,-1.5);
			
				\node[above] at (7.5,-1.3) {e};
					\node[above] at (9,-1.3) {f};
			%	\node[below] at (P) {$z_1$};
			%	\node[above] at (10.5,0.2) {$t$};
			%\path [->] (A)  edge[bend right=60]  node[above] {$0$} (D);
			%\path [->] (A)  edge[bend right=80]  node[above] {$0$} (E);
			\draw[draw = green] (4,0.5) -- (8,0.5);
			
			\draw[draw = green] (4,-0.5) -- (8,-0.5);
			\draw[draw = green] (4,0.5) -- (4,-0.5);
			\draw[draw = green] (8,0.5) -- (8,-0.5);
			\node (O) at (4.5,0.7) {{\color{green} $\bar{T}$}};
			\node (O) at (5,-1) {{\color{red} $\bar{H}$}};
			%\path [->] (B)  edge[bend left=80]  node[above] {$\beta_1$} (E);

			%	\path  (F) edge  (I);
			%	\path [->] (E)  edge[bend left=40]   (H);
			%	\path [->] (E)  edge[bend right=40]   (G);
			%	\path [->] (E) edge  (F);
			%	\path [->] (G)  edge[bend right=40]   (I);
			%	\path [->] (H)  edge[bend left=40]   (I);

		\end{tikzpicture}
	}
	\caption{Final configuration after moving holes $c$, $d$ and $g$ to $\bar{T}$.}
	\label{gather2}
	\end{subfigure}

\vspace{5mm}

\caption{Example of \textit{Gather hole} problem. We want to move three closest holes to the subtree $\bar{T}$.}
	\label{gather}
\end{figure}

	Figure \ref{gather} presents an example of the execution of the procedure just described. The blue circles represent the holes, and $\bar{H}=\{c,d,g\}$ is a subset of holes closest to subtree $\bar{T}$ (Figure \ref{gather1}). Another possible choice for $\bar{H}$ is, for example, $\bar{H}=\{c,e,g\}$. Figure \ref{gather2} shows the final configuration, in which the holes of $\bar{H}$ have been moved to the subtree.

	\section{Motion planning on trees}
\label{mpt}
Let $T=(V,E) $ be a tree, $P$ a set of pebbles and $H$ a set of holes. We assume that condition~(\ref{ass}) of Section~\ref{sec:assum} holds. Recall that
\begin{equation} 
	\label{cc}
c:=  \Bigg\{
	\begin{array}{ll}       	
		
		c_1, & \text{if $T$ is a path graph}  ,\\
		
		\max \{c_1+1,\, c_2 +2\}, & \text{otherwise}  ,\\

	\end{array}
\end{equation}
\noindent where $c_1$ is the maximum length of all the corridors and $c_2$ is the maximum length of the corridors with endpoints that are junctions.
Let  $\A^s$ be an initial valid configuration. Given a pebble $\bar{p}$ on $r=\A^s(\bar{p})$, and a target node $t\in V$, we show how to find a plan $f$ such that $t = \rho(\A^s,f)(\bar{p})$.
To do that, we analyze two cases:
\begin{itemize}
	\item[A.] $|\A^s(H) \cap T(F(r),t)| \geq c$, i.e., $T(F(r),t)$ contains at least $c$ holes.
	\item[B.] $|\A^s(H) \cap T(F(r),t) |< c$, i.e., $T(F(r),t)$ contains less than $c$ holes.
	\end{itemize}

For each of the two cases we present a solution procedure (\textit{Procedure A} and \textit{Procedure B}). The union of these procedures constitutes an algorithm to solve any instance of motion planning on trees, called the CATERPILLAR algorithm.

\vspace{1.5cm}
%\subsection{Case A}
\noindent \textbf{Case A.} $T(F(r),t)$ contains at least $c$ holes.\\

The main idea of the algorithm is to clear, piece by piece, the path that goes from $r$ to $t$, allowing the pebble to reach the target. We identify intermediate junctions on $\pi_{rt}$ (denoted by $i_k$), and "parking" positions (denoted by $\ell_{k}$) which are neighbor nodes of $i_k$, but do not belong to $\pi_{rt}$. The pebble moves from one parking position to the next one, until it reaches the target. When the pebble is on $\ell_k$, we move out of the way all the obstacles that are on $\pi_{i_k i_{k+1}}$, so that we can freely move the pebble from $\ell_{k}$ to $\ell_{k+1}$.
We identify a sequence of subsets of vertices on which the movement of the pebble from $\ell_{k}$ to $\ell_{k+1}$ will be defined: each of them will contain the path from $i_k $ to $i_{k+1}$ combined with the parking positions $\ell_{k}$ and $\ell_{k+1}$.
These subsets $(S_0,\dots ,S_m)$, called \textit{caterpillar sets}, are such that:

\begin{itemize}
	\item the restriction of $T$ to $S_k$ is a connected subtree  for all $k=0,\dots,m$;
	\item $|S_k|= c+1$ for all $k=0,\dots,m-1$, and $|S_m| \leq c+1$;
	\item $s \in S_0$ and $t \in S_m$;
	\item  $|S_k \cap S_{k+1}| \geq 2$ for all $k=0,\dots,m-1$, i.e., two consecutive sets have at least two nodes in common.
	\item $S_k \cap S_{k+1} \cap J \not = \emptyset$ for all $k=0,\dots,m-1$, i.e., two consecutive sets have at least one junction in common.
\end{itemize}

These properties guarantee that there are enough holes to clear the path and move the pebble from one parking position to the next one.
\subsubsection{Construction of caterpillar sets}
Along path $\pi_{rt}$ we select the node triple $(i_k,j_k,\ell_k)$: $i_k$ and $j_k$ 
represent the ends of a caterpillar set, while $\ell_k$ is a parking position. We proceed as follows:

\begin{enumerate}
	\item Let $\ell_0=r$, $i_0$ be the neighbor of $r$ that belongs to $\pi_{rt}$, and:
	\begin{itemize}
		\item if  $d(i_0,t) \leq c-1$, set $j_0 = t$, $m=0$ and stop;
		\item otherwise, let $j_0$ be the node on $\pi_{rt}$ such that $d(i_0,j_0)=c-2 $. Set $k=0$, $j_{-1}=i_0$ and go to Step 2.
	\end{itemize}
	
	\item let $i_{k+1}$ be the closest junction to $j_{k}$ on $\pi_{j_{k-1} j_{k}} \setminus \{j_{k-1}\}$, and  $\ell_{k+1}$ be one of the neighbors of $i_{k+1}$ not belonging to $\pi_{rt}$: 
	
	\begin{itemize}
		\item if  $d(i_{k+1},t)\leq c-1 $, set $j_{k+1} =t$, $m=k+1$ and stop;
		\item otherwise, let $j_{k+1}$ be the node on $\pi_{i_{k+1} t}$ such that $d(i_{k+1},j_{k+1})=c-2$. Set $k=k+1$ and repeat Step 2.
	\end{itemize}

	\end{enumerate}
\begin{figure}[h!]
	\centering
	\resizebox{0.8\columnwidth}{!}{
		\begin{tikzpicture}
			\node(A) at (0,0) {\textbullet};
			\node (B) at (1.5,0) {\textbullet};
		
			\node (J) at (1.5,1.5) {\textbullet};
			\node (C) at (3,0) {\textbullet};
			\node[below] at (A) {$r$};
				\node[above] at (A) {$\ell_0$};
					\node[below] at (B) {$i_0$};
					\node[above] at (C) {$i_1$};
			\node (K) at (3,-1.5) {\textbullet};
				\draw  (1.5,0) -- (1.5,1.5);
					\draw  (3,0) -- (3,-1.5);
					\node[left] at (K) {$\ell_1$};
			\node(D) at (4.5,0) {\textbullet};
		
			\node (E) at (6,0) {\textbullet};
				\node[above] at (E) {$j_0$};
			\node (F) at (7.5,0) {\textbullet} ;
				\node (L) at (7.5,-1.5) {\textbullet};
				\node[left] at (L) {$\ell_2$};
					\draw  (7.5,0) -- (7.5,-1.5);
			\node[above] at (F) {$j_1=i_2$};
				\node (G) at (9,0) {\textbullet} ;
					\draw  (9,0) -- (9,1.5);
						\node (M) at (9,1.5) {\textbullet};
				\node (H) at (10.5,0) {\textbullet} ;
					\node (I) at (12,0) {\textbullet} ;
					\node[above] at (I) {$t=j_2$};
					%	\draw  (12,0) -- (12,-1.5);
					%	\node (N) at (12,-1.5) {\textbullet};
			\draw  (0,0.05) -- (12,0.05);
			
			\node (O) at (0.2,1) {{\color{red}$S_0$}};
				\node (O2) at (6,1) {{\color{green}$S_1$}};
					\node (O) at (11,1) {{\color{blue}$S_2$}};
			
		\draw[draw = red] (-0.2,0.5) -- (6.3,0.5);
		\draw[draw = red] (-0.2,-2) -- (6.3,-2);
			\draw[draw = red] (-0.2,-2) -- (-0.2,0.5);
			\draw[draw = red] (6.3,0.5) -- (6.3,-2);
			
			\draw[draw = green] (2.5,0.8) -- (8.1,0.8);
			\draw[draw = green] (2.5,-1.7) -- (8.1,-1.7);
			\draw[draw = green] (2.5,0.8) -- (2.5,-1.7);
				\draw[draw = green] (8.1,0.8) -- (8.1,-1.7);
				
					\draw[draw = blue] (6.7,0.5) -- (12.5,0.5);
				\draw[draw = blue] (6.7,-2) -- (12.5,-2);
				\draw[draw = blue] (6.7,0.5) -- (6.7,-2);
			\draw[draw = blue] (12.5,0.5) -- (12.5,-2);
			%\path [->] (A)  edge[bend right=60]  node[above] {$0$} (D);
			%\path [->] (A)  edge[bend right=80]  node[above] {$0$} (E);

			%\path [->] (B)  edge[bend left=80]  node[above] {$\beta_1$} (E);

		%	\path  (F) edge  (I);
		%	\path [->] (E)  edge[bend left=40]   (H);
		%	\path [->] (E)  edge[bend right=40]   (G);
		%	\path [->] (E) edge  (F);
		%	\path [->] (G)  edge[bend right=40]   (I);
		%	\path [->] (H)  edge[bend left=40]   (I);

		\end{tikzpicture}
	}
	\caption{We consider the motion planning problem with source vertex $r$ and target vertex $t$ on a tree with $c=5$. $S_0$, $S_1$ and $S_2$ are the \textit{caterpillar sets} along path $\pi_{rt}$.}
	\label{caterpillar}
\end{figure}
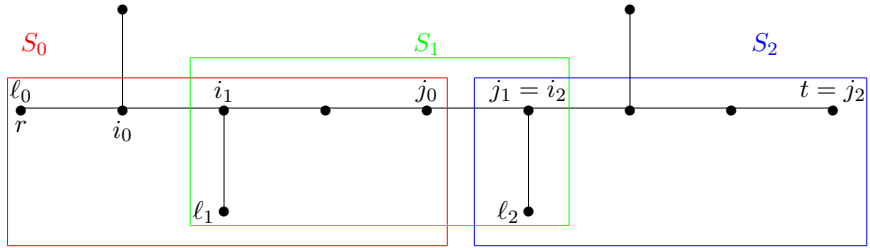

\begin{rem}
	Note that $i_{k+1}$ is always well defined. Indeed, $d(i_k,j_k)=c-2\geq c_2$ guarantees that there exists a junction on $\pi_{i_k j_k} \setminus \{i_k\}$ ($i_{k+1} \in \pi_{i_k j_k} \setminus \{i_k\}$). Since, by definition, 
$i_k$ is the closest junction to $j_{k-1}$ on $\pi_{j_{k-2} j_{k-1}} \setminus \{j_{k-2}\}$, then $i_{k+1} \not \in \pi_{i_{k}j_{k-1}}$. Therefore, there exists $i_{k+1} \in \pi_{j_{k-1}j_k}\setminus \{j_{k-1}\}$ which is the closest junction to $j_k$.
\end{rem}

\begin{obser}
	
	Since $d(i_{k},j_{k})=c-2$, it follows that 
		\[ d(i_{k},i_{k+2}) =  d(i_{k},j_{k}) +  d(j_{k},i_{k+2})\geq c-1. \] 
	Now, let us find a lower bound for 
	\[ d(i_0,i_m) = \sum_{k=0}^{m-1}d(i_k,i_{k+1}) .\]
	If $m$ is even,
	\[ \sum_{k=0}^{m-1}d(i_k,i_{k+1})  = \sum_{k=0}^{\frac{m}{2}-1} [d(i_{2k},i_{2k+1} )+ d(i_{2k+1},i_{2k+2}) ]=\]
	
	\[=   \sum_{k=0}^{\frac{m}{2}-1} d(i_{2k},i_{2k+2}) \geq \frac{m}{2} \cdot (c-1). \]

Otherwise, if $m$ is odd,
	\[ \sum_{k=0}^{m-1}d(i_k,i_{k+1})  = \left(\sum_{k=0}^{m-2}d(i_k,i_{k+1})\right) + d(i_{m-1},i_{m}) \]
	
	\[ \geq  \frac{m-1}{2} \cdot (c-1).  \]

	Therefore,
\[m \leq 2 \cdot \frac{d(i_0,i_m)}{c-1} +1 \leq 2 \cdot \frac{\delta}{c-1} +1, \]

where $\delta$ is the diameter of the tree. If $c \geq 2$, it follows that $m=O( \frac{\delta}{c})$. Note that $c=1$ only holds for the trivial case of the tree with 2 edges.\\

\end{obser}
Nodes $(i_k,j_k,\ell_k)$ are used to delimit the borders of the caterpillar sets (see Figure \ref{caterpillar}), which are  defined as follows:

 \[S_k = \pi_{i_k j_k} \cup \{\ell_k\}\cup \{\ell_{k+1}\}, \quad \forall k=0,\dots,m-1,\]
 
\[S_m = \pi_{i_m j_m} \cup \{\ell_m\}.\]

 We can easily note that the union of all caterpillar sets is a caterpillar tree (i.e., a tree in which all the vertices are within distance 1 from a central path). Moreover, all the properties of the caterpillar sets are verified:

\begin{itemize}
\item $S_k$ is a connected component of $T$, and so it is a subtree;

\item  $|S_k| = |\pi_{i_k j_k}| +2 = c+1$ for all $k=0,\dots,m-1$ and $|S_m| = |\pi_{i_{m} j_m}| +2 \leq c+1$;

\item $\{i_{k+1},l_{k+1}\} \subset S_k \cap S_{k+1} $ and $i_{k+1} \in J $,  so $|S_k \cap S_{k+1}| \geq 2$ and $S_k \cap S_{k+1} \cap J \not = \emptyset$ for $k=0,\dots,m-1$.

 \end{itemize}

 Moreover, it holds that

\[|S_{k+1} \cup S_{k}| = |S_{k+1}| + |S_{k}| - |S_{k+1} \cap S_{k}|\leq\]
\[ \leq (2c +2) -2 = 2c.\]

We are now ready to describe the procedure for solving the motion planning problem in case A.  \\

%\vspace{1cm}

\begin{figure}[h!]
	\centering
		\begin{subfigure}[t]{0.8\columnwidth}
		\centering
		\resizebox{\columnwidth}{!}{
	\begin{tikzpicture}
		[scale=1.2,auto=left]
		\draw[dashed,draw = orange] (-0.3,0.5) -- (3.4,0.5);
			\draw[dashed,draw = orange] (-0.3,-2) -- (3.4,-2);
				\draw[dashed,draw = orange] (-0.3,-2) -- (-0.3,0.5);
					\draw[dashed,draw = orange] (3.4,0.5) -- (3.4,-2);
					
			\draw[dashed,draw = cyan] (2.6,0.6) -- (6.3,0.6);
			\draw[dashed,draw = cyan] (2.6,-1.9) -- (6.3,-1.9);
			\draw[dashed,draw = cyan] (2.6,0.6) -- (2.6,-1.9);
			\draw[dashed,draw = cyan] (6.3,0.6) -- (6.3,-1.9);

			\node (q) at (0.2,0.7) {{\color{orange}$S_k$}};
			\node (q) at (5.7,0.8) {{\color{cyan}$S_{k+1}$}};

		\node[style={circle,fill=blue!30}] (A) at (0,0) {\textbullet};
		\node[style={circle,fill=blue!30}] (B) at (1.5,0) {\textbullet};
		\node[below] at (0.3,-0.2) {$i_k$};
		\node[style={circle,fill=blue!30}] (J) at (0,-1.5) {\textbullet};
		%\node[below] at (1.5,-0.2) {$u_4$};
		\node[style={circle,fill=blue!30}] (C) at (3,0) {\textbullet};
		\node[below] at (3.3,-0.2) {$j_k \equiv i_{k+1}$};
		%	\node[below] at (A) {$s$};
		%	\node[above] at (A) {$f_0$};
		%	\node[below] at (B) {$i_0$};
		%	\node[above] at (C) {$i_1$};
		\node[style={rectangle,fill=red!40}] (K) at (3,-1.5) {\textbullet};

		%\node[left] at (K) {$f_1$};
		\node[style={rectangle,fill=green!40}] (D) at (4.5,0) {\textbullet};
			\node[style={rectangle,fill=green!40}] (F) at (6,-1.5) {\textbullet};
		%	\node[above] at (D) {$z_2$};
		\node[style={rectangle,fill=green!40}] (E) at (6,0) {\textbullet};
		%	\node[below] at (6.15,0) {$s$};
		\node[right] at (3.2,-1.5) {$\ell_{k+1}$};
			\node[right] at (0.2,-1.5) {$\ell_{k}$};
				\node[right] at (6.2,-1.5) {$\ell_{k+2}$};
		\node[below] at (6,-0.2) {$j_{k+1}$};
		
			\draw  (0,0) -- (0,-1.5);
		\draw  (6,0) -- (6,-1.5);
		\draw  (3,0) -- (3,-1.5);
		
		\draw  (0,0.05) -- (6,0.05);
			\path[->]  (5,0.8)  edge[draw=red]  node[above]{{\color{red}(1) slide obstacles}}  (1,0.8);
		\path[->]  (K)  edge[bend left=80,draw=red]    (F);
					\node (s) at (4.5,-1) {{\color{red}(2) move  pebble }};
			%	\path[->]  (E)  edge[bend right=50,draw=red]  node[above]{{\color{red}$(2)$}}  (A);
		%\path[->]  (F)  edge[bend right=80,draw=red]  node[above]{{\color{red}$(3)$}}  (B);
		
	\end{tikzpicture}
}
\caption{Initial configuration. First we have slide the obstacles from $S_{k+1} \setminus S_{k}$ to $S_{k} \setminus S_{k+1} $, then we move the pebble from  $\ell_{k+1}$ to  $\ell_{k+2}$. }
\end{subfigure}

\vspace{6mm}

		\begin{subfigure}[t]{0.8\columnwidth}
	\centering
	\resizebox{\columnwidth}{!}{
		\begin{tikzpicture}
			[scale=1.2,auto=left]
			\draw[dashed,draw = orange] (-0.3,0.5) -- (3.4,0.5);
			\draw[dashed,draw = orange] (-0.3,-2) -- (3.4,-2);
			\draw[dashed,draw = orange] (-0.3,-2) -- (-0.3,0.5);
			\draw[dashed,draw = orange] (3.4,0.5) -- (3.4,-2);
			
			\draw[dashed,draw = cyan] (2.6,0.6) -- (6.3,0.6);
			\draw[dashed,draw = cyan] (2.6,-1.9) -- (6.3,-1.9);
			\draw[dashed,draw = cyan] (2.6,0.6) -- (2.6,-1.9);
			\draw[dashed,draw = cyan] (6.3,0.6) -- (6.3,-1.9);

				\node (q) at (0.2,0.7) {{\color{orange}$S_k$}};
					\node (q) at (5.7,0.8) {{\color{cyan}$S_{k+1}$}};
			
			\node[style={rectangle,fill=green!40}] (A) at (0,0) {\textbullet};
			\node[style={rectangle,fill=green!40}] (B) at (1.5,0) {\textbullet};
			\node[below] at (0.3,-0.2) {$i_k$};
			\node[style={rectangle,fill=green!40}] (J) at (0,-1.5) {\textbullet};
			%\node[below] at (1.5,-0.2) {$u_4$};
			\node[style={circle,fill=blue!30}] (C) at (3,0) {\textbullet};
			\node[below] at (3.3,-0.2) {$j_k \equiv i_{k+1}$};
			%	\node[below] at (A) {$s$};
			%	\node[above] at (A) {$f_0$};
			%	\node[below] at (B) {$i_0$};
			%	\node[above] at (C) {$i_1$};
			\node[style={circle,fill=blue!30}] (K) at (3,-1.5) {\textbullet};

			%\node[left] at (K) {$f_1$};
			\node[style={circle,fill=blue!30}] (D) at (4.5,0) {\textbullet};
			\node[style={rectangle,fill=red!40}] (F) at (6,-1.5) {\textbullet};
			%	\node[above] at (D) {$z_2$};
			\node[style={circle,fill=blue!30}] (E) at (6,0) {\textbullet};
			%	\node[below] at (6.15,0) {$s$};
			\node[right] at (3.2,-1.5) {$\ell_{k+1}$};
			\node[right] at (0.2,-1.5) {$\ell_{k}$};
			\node[right] at (6.2,-1.5) {$\ell_{k+2}$};
			\node[below] at (6,-0.2) {$j_{k+1}$};
			
			\draw  (0,0) -- (0,-1.5);
			\draw  (6,0) -- (6,-1.5);
			\draw  (3,0) -- (3,-1.5);
			
			\draw  (0,0.05) -- (6,0.05);
			
			%	\path[->]  (E)  edge[bend right=50,draw=red]  node[above]{{\color{red}$(2)$}}  (A);
			%\path[->]  (F)  edge[bend right=80,draw=red]  node[above]{{\color{red}$(3)$}}  (B);
			
		\end{tikzpicture}
	
	}

\caption{Final configuration.}
\end{subfigure}
\vspace{4mm}

\caption{Example of execution of an iteration of the {\tt for} cycle of Step 3 of \textit{Procedure A}. Blue circles are holes, green squares are obstacles and the red square is the marked pebble.}
\label{procA}
\end{figure}
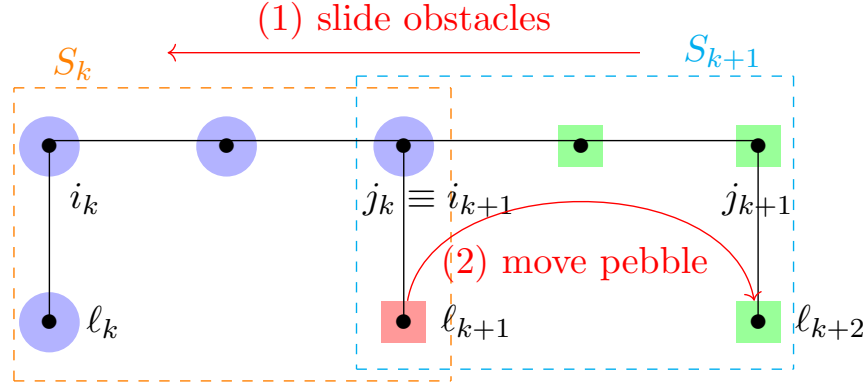
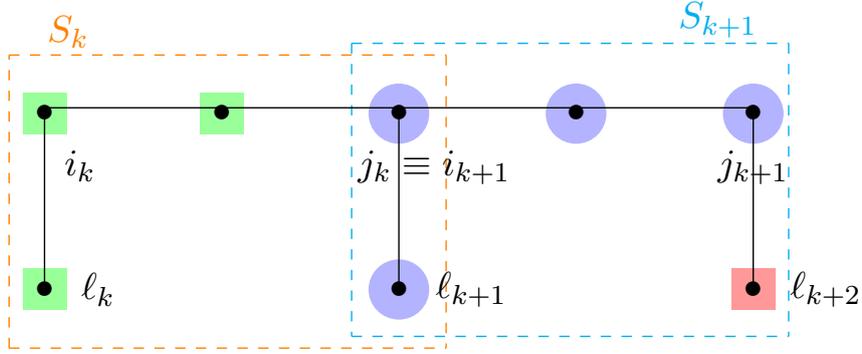

\noindent \textit{Procedure A}
\begin{enumerate}
	\item Gather holes in $S_0 \setminus \{\ell_0\} $ : $O(n c)$ moves in view of Proposition \ref{prop:gatherhole}.
	
	\item Move pebble from $\ell_0$ to $\ell_1$: $c$ moves.
	
	\item For all $k=0,..,m-1$ moves the holes from $S_k \setminus \{\ell_{k+1}\}$ to  $S_{k+1} \setminus \{\ell_{k+1}\}$ and move pebble $p$ from $\ell_{k+1}$ to $t$ or $\ell_{k+2}$ (see Figure \ref{procA}). The former operation consists in  sliding the obstacles from $S_{k+1} \setminus S_{k}$ to $S_{k} \setminus S_{k+1} $: this is equivalent to \textit{ gather holes in $S_{k+1}$} in the subtree obtained by the restriction of $T$ to $S_{k+1} \cup S_{k}$ , and so it requires at most $(|S_{k+1} \cup S_{k}| \cdot c)=O(c^2)$ moves.
	The last operation  requires at most $c$ moves. Overall the {\tt for} cycle has length complexity $O(m c^2)$.
\end{enumerate}
Note that the operations just described in \textit{Procedure A} are all feasible because the properties of caterpillar sets hold.

Then, we proved the following result.

\begin{prop}
\label{prop:procA}
The length complexity of the solution provided by Procedure A is $O(n c)$.
\end{prop}
\begin{proof}
	The solution provided by Procedure A requires
	\[O(n  c) + O(m c^2)\]
    moves. Equivalently, the length complexity of the solution is
	\[O(n  c) + O(\delta c ) = O(n c),\]
	recalling that $m=O\left( \frac{\delta}{c}\right)$.\\
\end{proof}

\noindent \textbf{Case B.} $T(F(s),t)$ contains less than $c$ holes.\\

Suppose that $q$ holes are missing to get to $c$.
In this case, we create a neighborhood of $s$ made up of $q$ holes in $C(F(s),t)$, and then we move the pebble $p$ to a node $v$ such that $T(F(v),t)$ contains at least $c$ holes.
Let $z_1,\dots,z_k$ the neighbors of $s$ which are not in $T(F(s),t)$. For all $j=1,\dots,k$, we consider $T_j := T(F(s), z_j)$, the subtree containing $z_j$, and $V_j$ the corresponding set of nodes. 
In this case the motion planning problem is solved through the following procedure.\\

\noindent \textit{Procedure B}

\begin{enumerate}
	\item Set $j=1$;
	\item Let $q_j = |\A^s(H) \cap V_j|$  be the number of holes in $T_j$:
	\begin{itemize}
		\item if $q_j \leq q$, gather all the holes of $T_j$ in $H_j$, which is a subset of $V_j$ of cardinality $q_j$  closest to the subset $\{s\}$:
		
		\[
		H_j \in \arg \min_{\begin{array}{c} \scriptstyle W \in \PP(V_j) \\ 	
				\scriptstyle 	|W| = q_j\end{array}
		} d(W,\{s\});
		\]
		 Set $j=j+1$, $q=q-q_j$ and go back to Step 2;
		 
		\item otherwise, gather $q$ holes of $T_j$ in $H_j$ which, in this case, is defined as a subset of $V_j$ of cardinality $q$  closest to $s$.\\
		
	\end{itemize}

\item choose $v\in H_j $ which has the maximum distance from $s$ and move pebble $p$ on $v$;
\item  reinitialize $s$ with node $v$ and apply \textit{Procedure A}.\\
\end{enumerate}

Note that Step 4 is feasible because in $T(F(v),t)$ there are certainly at least $c$ holes.

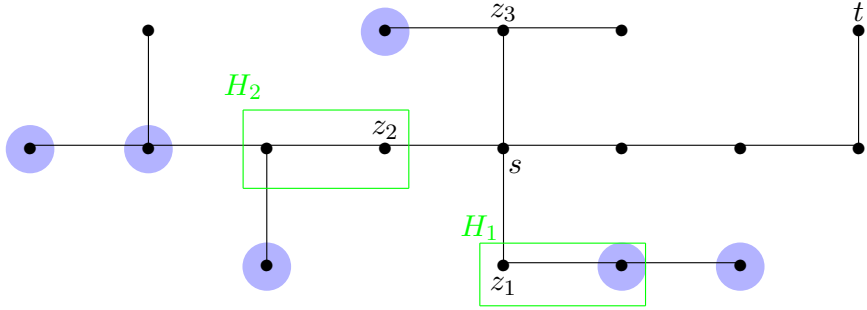
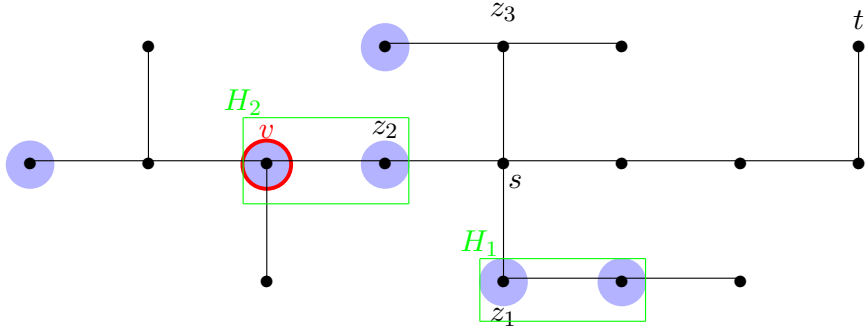
\begin{figure}[h!]
	\centering
	\begin{subfigure}[t]{\columnwidth}
		\centering
	\resizebox{0.8\columnwidth}{!}{
		\begin{tikzpicture}
			\node[style={circle,fill=blue!30}] (A) at (0,0) {\textbullet};
			\node[style={circle,fill=blue!30}] (B) at (1.5,0) {\textbullet};
			
			\node (J) at (1.5,1.5) {\textbullet};
			\node (C) at (3,0) {\textbullet};
		%	\node[below] at (A) {$s$};
		%	\node[above] at (A) {$f_0$};
		%	\node[below] at (B) {$i_0$};
		%	\node[above] at (C) {$i_1$};
			\node[style={circle,fill=blue!30}] (K) at (3,-1.5) {\textbullet};
			
			\draw  (1.5,0) -- (1.5,1.5);
			\draw  (3,0) -- (3,-1.5);
			%\node[left] at (K) {$f_1$};
			\node(D) at (4.5,0) {\textbullet};
			\node[above] at (D) {$z_2$};
			\node (E) at (6,0) {\textbullet};
			\node[below] at (6.15,0) {$s$};
			\node (F) at (7.5,0) {\textbullet} ;
			
			\node[style={circle,fill=blue!30}]   (L) at (7.5,-1.5) {\textbullet};
		%	\node[left] at (L) {$f_2$};
			%\draw (7.5,0) -- (7.5,-1.5);
		%	\node[above] at (F) {$j_1=i_2$};
			\node (G) at (9,0) {\textbullet} ;
			\draw  (10.5,0) -- (10.5,1.5);
			\node (M) at (10.5,1.5) {\textbullet};
			\node (H) at (10.5,0) {\textbullet} ;
			%\node (I) at (12,0) {\textbullet} ;
			%\node[above] at (I) {$t=j_2$};
			%	\draw  (12,0) -- (12,-1.5);
			%	\node (N) at (12,-1.5) {\textbullet};
				\node (I) at (6,1.5) {\textbullet};
					\node[above] at (I) {$z_3$};
					\node[style={circle,fill=blue!30}]  (N) at (4.5,1.5) {\textbullet};
						\node (O) at (7.5,1.5) {\textbullet};
			\draw  (0,0.05) -- (10.5,0.05);
				\draw  (4.5,1.55) -- (7.5,1.55);
			\draw  (6,0) -- (6,1.5);
			
			\node (P) at (6,-1.5) {\textbullet};
			\node[style={circle,fill=blue!30}] (Q) at (9,-1.5) {\textbullet};
			\draw (6,-1.45) -- (9,-1.45);
		\draw (6,0) -- (6,-1.5);
			\node[below] at (P) {$z_1$};
				\node[above] at (10.5,1.5) {$t$};
				
					\draw[draw = green] (2.7,0.5) -- (4.8,0.5);
				
				\draw[draw = green] (2.7,-0.5) -- (4.8,-0.5);
				\draw[draw = green] (2.7,0.5) -- (2.7,-0.5);
				\draw[draw = green] (4.8,0.5) -- (4.8,-0.5);
					\node at (2.7,0.8) {{\color{green}$H_2$}};
				\draw[draw = green] (5.7,-2) -- (7.8,-2);			
				\draw[draw = green] (5.7,-1.2) -- (7.8,-1.2);
				\draw[draw = green](5.7,-2) --  (5.7,-1.2);
				\draw[draw = green] (7.8,-2) -- (7.8,-1.2);
				
				\node at (5.7,-1) {{\color{green}$H_1$}};
			%\path [->] (A)  edge[bend right=60]  node[above] {$0$} (D);
			%\path [->] (A)  edge[bend right=80]  node[above] {$0$} (E);

			%\path [->] (B)  edge[bend left=80]  node[above] {$\beta_1$} (E);

			%	\path  (F) edge  (I);
			%	\path [->] (E)  edge[bend left=40]   (H);
			%	\path [->] (E)  edge[bend right=40]   (G);
			%	\path [->] (E) edge  (F);
			%	\path [->] (G)  edge[bend right=40]   (I);
			%	\path [->] (H)  edge[bend left=40]   (I);

		\end{tikzpicture}
	}
	\caption{Initial configuration. }
	\label{complexity_ex1}
\end{subfigure}

\vspace{5mm}

\begin{subfigure}[t]{0.8\columnwidth}
\centering
	\centering
	\resizebox{\columnwidth}{!}{
		\begin{tikzpicture}
			\node[style={circle,fill=blue!30}] (A) at (0,0) {\textbullet};
			\node (B) at (1.5,0) {\textbullet};
			
			\node (J) at (1.5,1.5) {\textbullet};
			\node[style={circle,fill=blue!30}, shape=circle,draw=red,line width=0.5mm] (C) at (3,0) {\textbullet};
			\node[above] at (3,0.2) {{\color{red}$v$}};
			%	\node[below] at (A) {$s$};
			%	\node[above] at (A) {$f_0$};
			%	\node[below] at (B) {$i_0$};
			%	\node[above] at (C) {$i_1$};
			\node (K) at (3,-1.5) {\textbullet};
			
			\draw  (1.5,0) -- (1.5,1.5);
			\draw  (3,0) -- (3,-1.5);
			%\node[left] at (K) {$f_1$};
			\node[style={circle,fill=blue!30}] (D) at (4.5,0) {\textbullet};
			\node[above] at (4.5,0.2) {$z_2$};
			\node (E) at (6,0) {\textbullet};
			\node[below] at (6.15,0) {$s$};
			\node (F) at (7.5,0) {\textbullet} ;
		%	\node[above] at (F) {$z_1$};
			\node[style={circle,fill=blue!30}]   (L) at (7.5,-1.5) {\textbullet};
			%	\node[left] at (L) {$f_2$};
		%	\draw (7.5,0) -- (7.5,-1.5);
			%	\node[above] at (F) {$j_1=i_2$};
			\node (G) at (9,0) {\textbullet} ;
			\draw  (10.5,0) -- (10.5,1.5);
			\node (M) at (10.5,1.5) {\textbullet};
			\node  (H) at (10.5,0) {\textbullet} ;
			%\node (I) at (12,0) {\textbullet} ;
			%\node[above] at (I) {$t=j_2$};
			%	\draw  (12,0) -- (12,-1.5);
			%	\node (N) at (12,-1.5) {\textbullet};
			\node (I) at (6,1.5) {\textbullet};
			\node[above] at (6,1.7) {$z_3$};
			\node[style={circle,fill=blue!30}]  (N) at (4.5,1.5) {\textbullet};
			\node (O) at (7.5,1.5) {\textbullet};
			\draw  (0,0.05) -- (10.5,0.05);
			\draw  (4.5,1.55) -- (7.5,1.55);
			\draw  (6,0) -- (6,1.5);
			
				\node[style={circle,fill=blue!30}] (P) at (6,-1.5) {\textbullet};
			\node (Q) at (9,-1.5) {\textbullet};
			\draw (6,-1.45) -- (9,-1.45);
			\draw (6,0) -- (6,-1.5);
			\node[below] at (6,-1.7) {$z_1$};
			\node[above] at (10.5,1.6) {$t$};

				\draw[draw = green] (2.7,0.6) -- (4.8,0.6);
			
			\draw[draw = green] (2.7,-0.5) -- (4.8,-0.5);
			\draw[draw = green] (2.7,0.6) -- (2.7,-0.5);
			\draw[draw = green] (4.8,0.6) -- (4.8,-0.5);
			
			\node at (2.7,0.8) {{\color{green}$H_2$}};

				\draw[draw = green] (5.7,-2) -- (7.8,-2);			
			\draw[draw = green] (5.7,-1.2) -- (7.8,-1.2);
			\draw[draw = green](5.7,-2) --  (5.7,-1.2);
			\draw[draw = green] (7.8,-2) -- (7.8,-1.2);
			
			\node at (5.7,-1) {{\color{green}$H_1$}};
			%\path [->] (A)  edge[bend right=60]  node[above] {$0$} (D);
			%\path [->] (A)  edge[bend right=80]  node[above] {$0$} (E);

			%\path [->] (B)  edge[bend left=80]  node[above] {$\beta_1$} (E);

			%	\path  (F) edge  (I);
			%	\path [->] (E)  edge[bend left=40]   (H);
			%	\path [->] (E)  edge[bend right=40]   (G);
			%	\path [->] (E) edge  (F);
			%	\path [->] (G)  edge[bend right=40]   (I);
			%	\path [->] (H)  edge[bend left=40]   (I);

		\end{tikzpicture}
	}
	\caption{Final configuration after \textit{Procedure B}.}
	\label{complexity_ex2}
	\end{subfigure}

\vspace{5mm}

	\caption{Example of situation of Case B with $q=4$ missing holes.}
\end{figure}
\vspace{5mm}

Let $n_j$ be the number of nodes of $T_j$. Then, recalling the length complexity of the gather holes procedure, \textit{Procedure B} (without the final application of \textit{Procedure A}) requires at most $\left(\sum_{j=1}^k n_j  q_j\right)$ moves to bring the holes in the neighborhood of $s$, and at most $n$ moves to bring pebble $p$ on $v$.
Therefore, the length complexity of this procedure is $O(n q)$ which, in the worst case, is $O(n c)$. The final application of \textit{Procedure A} does not modify the complexity result.
Thus, we proved the following result.
\begin{prop}
\label{prop:procB}
The complexity of Procedure B is $O(nc)$.\\
\end{prop}
\section{Pebble motion on trees}
\label{sec:pmt}
Now we are ready to provide a procedure for the solution of PMT.
	We are given a tree $T=(V,E)$, a pebble set $P$, an initial valid configuration $\A^s$, and a final valid configuration $\A^t$. As already mentioned, the PMT problem consists in generating a plan $f$ such that $\A^t(p) = \rho(\A^s,f)(p)$ $\forall \, p \in P$.
        We will use the following strategy. First, we solve an unlabeled PMT on $T$ to bring the pebbles to an ordered set of $\textit{ intermediate targets}$, $\bar{t}_1,\ldots,\bar{t}_{|P|}$. Later, we will discuss the choice of these targets. We call $g$ the corresponding plan. Note that we do not apply this plan, but will apply the inverse plan $g^{-1}$ at the end of the procedure.
Then, we solve a set of $|P|$ motion planning problems on a sequence $T_k$ of subtrees of $T$. Over each subtree, we use the CATERPILLAR algorithm to move each pebble $p \in P$ to one intermediate target $t_k$ (that occupies a leaf of $T_k$). After the execution of the corresponding plan, we remove target $v$ from tree $T_k$, obtaining tree $T_{k+1}$, and remove $p$ from the set of pebbles. Finally, we apply the inverse plan $g^{-1}$ on $T$.
	
%To do that, first we define a set of \textit{ intermediate targets}.
We choose the \textit{intermediate targets} such that trees $T_k$ satisfies the following properties, for $k=1,\ldots,|P|-1$:
\begin{itemize}
\item $T_k$ contains $\bar{t}_k,\bar{t}_{k+1},\ldots,\bar{t}_{|P|}$ but does not contain $\bar{t}_1,\ldots,\bar{t}_{k-1}$;
  \item $c(T_k) \geq c(T_{k+1})$.
\end{itemize}

Let $L(T_k)$ be the set of all the leaves of the tree $T_k$. 
 We denote by $V_1= \{\bar{t}_k\, : k\in \{1, \dots, |P|\}\}$ the set of intermediate targets. We define them by the following procedure:

\begin{enumerate}
	\item Set $k=1$.
	\item If $k>|P|$, stop.
	\item Select $v\in L(T_k)$ such that, given the tree  $T_k^v$ obtained by removing $v$ from $T_k$, it holds that $c(T_k^v)\leq c(T_k)$.
	
	\item Define $\bar{t}_k=v$ and $T_{k+1}=T_k^v$. Set $k = k+1$ and to go Step 2.
\end{enumerate}

\vspace{0.5cm}

\begin{figure}[htp!]
	\centering
	\begin{subfigure}[t]{0.23\columnwidth}
		\centering
		\resizebox{\columnwidth}{!}{
			\begin{tikzpicture}
			
				%\node[left] at (K) {$f_1$};
				\node(D)at (4.5,0) {\textbullet};
				\node[above] at (D) {{$v$}};
				\node  (E) at (6,0) {\textbullet};
					\node[above] at (6.25,0) {$n_v$};
				\node (F) at (7.5,0) {\textbullet} ;
				%	\node[right] at (F) {$t_2$};
				
				\node   (L) at (7.5,-1.5) {\textbullet};
				%	\node[right] at (L) {$t_1$};
				%	\node[left] at (L) {$f_2$};
				%\draw (7.5,0) -- (7.5,-1.5);
				%	\node[above] at (F) {$j_1=i_2$};
				
				\node (I) at (6,1.5) {\textbullet};
				
			%	\node[above] at (I) {{\color{red}$t$}};
				\node  (N) at (4.5,1.5) {\textbullet};
				\node (O) at (7.5,1.5) {\textbullet};
				%	\node[right] at (O) {$t_3$};
				\draw  (4.5,0.05) -- (7.5,0.05);
				\draw  (4.5,1.55) -- (7.5,1.55);
				\draw  (6,0) -- (6,1.5);
				
				\node (P) at (6,-1.5) {\textbullet};
				
				\draw (6,-1.45) -- (7.5,-1.45); 	
				\draw (6,0) -- (6,-1.5);

			\end{tikzpicture}
		}
		\caption{Case $v \in L_4(T_k)$. }
		\label{fig:inttarget1}
	\end{subfigure}
	\begin{subfigure}[t]{0.23\columnwidth}
		\centering
		\resizebox{\columnwidth}{!}{
			\begin{tikzpicture}
				
				%\node[left] at (K) {$f_1$};
			
				\node  (E) at (6,0) {\textbullet};
			
				\node (F) at (7.5,0) {\textbullet} ;
				%	\node[right] at (F) {$t_2$};
				
				\node   (L) at (7.5,-1.5) {\textbullet};
					\node[above] at (L) {$v$};
				%	\node[right] at (L) {$t_1$};
				%	\node[left] at (L) {$f_2$};
				%\draw (7.5,0) -- (7.5,-1.5);
				%	\node[above] at (F) {$j_1=i_2$};
				
				\node (I) at (6,1.5) {\textbullet};
				
				%	\node[above] at (I) {{\color{red}$t$}};
				\node  (N) at (4.5,1.5) {\textbullet};
				\node (O) at (7.5,1.5) {\textbullet};
				%	\node[right] at (O) {$t_3$};
				\draw  (6,0.05) -- (7.5,0.05);
				\draw  (4.5,1.55) -- (7.5,1.55);
				\draw  (6,0) -- (6,1.5);
				
				\node (P) at (6,-1.5) {\textbullet};
					\node[left] at (P) {$n_v$};
				\draw (6,-1.45) -- (7.5,-1.45); 	
				\draw (6,0) -- (6,-1.5);

			\end{tikzpicture}
		}
		\caption{Case $v \in L_2(T_k)$. }
		\label{fig:inttarget2}
	\end{subfigure}
	\begin{subfigure}[t]{0.23\columnwidth}
	\centering
	\resizebox{\columnwidth}{!}{
		\begin{tikzpicture}
			
			%\node[left] at (K) {$f_1$};
			
			\node  (E) at (6,0) {\textbullet};
			\node[left] at (E) {$n_v$};
			\node (F) at (7.5,0) {\textbullet} ;
				\node[above] at (F) {$v$};
			%	\node[right] at (F) {$t_2$};

			%	\node[right] at (L) {$t_1$};
			%	\node[left] at (L) {$f_2$};
			%\draw (7.5,0) -- (7.5,-1.5);
			%	\node[above] at (F) {$j_1=i_2$};
			
			\node (I) at (6,1.5) {\textbullet};
			
			%	\node[above] at (I) {{\color{red}$t$}};
			\node  (N) at (4.5,1.5) {\textbullet};
			\node (O) at (7.5,1.5) {\textbullet};
			%	\node[right] at (O) {$t_3$};
			\draw  (6,0.05) -- (7.5,0.05);
			\draw  (4.5,1.55) -- (7.5,1.55);
			\draw  (6,0) -- (6,1.5);
			
			\node (P) at (6,-1.5) {\textbullet};

			\draw (6,0) -- (6,-1.5);

		\end{tikzpicture}
}
	\caption{Case $v \in L_3(T_k)$, $\bar{C}(T_k)\not = \emptyset$. }
	\label{fig:inttarget3}
\end{subfigure}
\begin{subfigure}[t]{0.23\columnwidth}
	\centering
	\resizebox{\columnwidth}{!}{
		\begin{tikzpicture}
			
			%\node[left] at (K) {$f_1$};
			
			\node  (E) at (6,0) {\textbullet};

			%	\node[right] at (F) {$t_2$};

			%	\node[right] at (L) {$t_1$};
			%	\node[left] at (L) {$f_2$};
			%\draw (7.5,0) -- (7.5,-1.5);
			%	\node[above] at (F) {$j_1=i_2$};
			
			\node (I) at (6,1.5) {\textbullet};
			\node[above] at (I) {$n_v$};
			%	\node[above] at (I) {{\color{red}$t$}};
			\node  (N) at (4.5,1.5) {\textbullet};
			\node (O) at (7.5,1.5) {\textbullet};
				\node[above] at (O) {$v$};
			%	\node[right] at (O) {$t_3$};
		
			\draw  (4.5,1.55) -- (7.5,1.55);
			\draw  (6,0) -- (6,1.5);

		\end{tikzpicture}
	}
	\caption{Case $v \in L_3(T_k)$, $\bar{C}(T_k) = \emptyset$. }
	\label{fig:inttarget4}
\end{subfigure}

\caption{\label{fig:fourcases} The four cases of Proposition \ref{prop:leaves}.}
\end{figure}
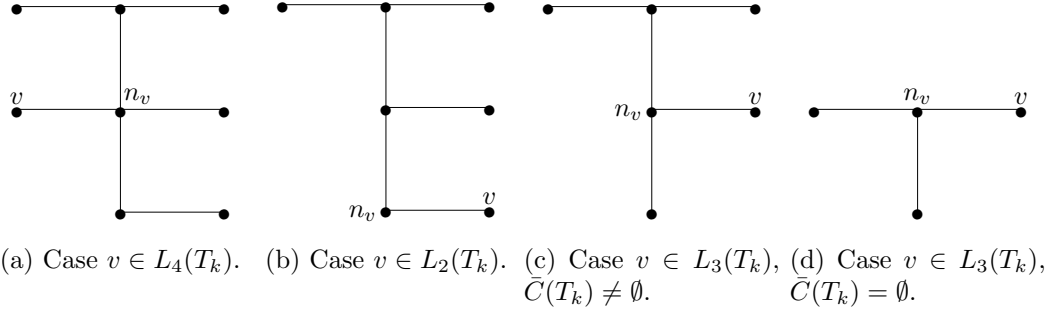

The following proposition shows that Step 3 of the above procedure is well defined, i.e., it is always possible to find a leaf $v$ of $T_k$ such that
$c(T_k^v)\leq c(T_k)$. Note that the result is valid for general trees and not only for the subtrees $T_k$ generated by the above procedure.
\begin{prop}
	\label{prop:leaves}
	For all  $k\in \{1, \dots, |P|-1\}$, there exists $v \in L(T_{k}) $ such that $c(T_{k}^v) \leq c(T_k)$, where $T_{k}^v$ is the tree obtained by removing $v$ from $T_{k}$.
\end{prop}
\begin{proof}
	For each $w\in L(T_{k})$ we denote by $n_w$ the unique neighbor of $w$, and by $deg(n_w)$ its degree. We define three subsets of leaves:
	$L_2(T_{k}):= \{w \in L(T_{k}) \, : \, deg(n_w) =2\}$, 	$L_3(T_{k}):= \{w \in L(T_{k}) \, : \, deg(n_w) =3\}$, 	$L_4(T_{k})= \{w \in L(T_{k}) \, : \, deg(n_w) > 3\}$. Note that $ L_2(T_{k}) \cup L_3(T_{k}) \cup L_4(T_{k}) = L(T_{k})$. We choose $v$ as follows  (see also Figure \ref{fig:fourcases}):
	\begin{itemize}
		\item if $L_4(T_{k}) \not = \emptyset$, we choose $v \in L_4(T_{k})$: in this case $c(T_{k}^v)= c(T_k)$, since by removing $v$, we just remove a corridor of length 1 (see Figure \ref{fig:inttarget1});
		\item else, if $L_2(T_{k}) \not = \emptyset$, we choose $v\in L_2(T_{k})$: in this case $c_1(T_{k}^v) \leq c_1(T_k)$ and then $c(T_{k}^v) \leq c(T_k)$, since by removing $v$, we are reducing by one the length of one corridor in $C(T_k)\setminus \bar{C}(T_k)$ (see Figure \ref{fig:inttarget2});
		\item otherwise,
all corridors in $C(T_k)\setminus \bar{C}(T_k)$ have length one. We choose $v \in L_3(T_k)$.
Two cases are possible:
\begin{enumerate}
%Therefore, it holds that $c_1(T_{k})= c_2(T_{k})$ and, consequently, $c(T_k)=c_2(T_k)+2$. Now, there are two possibilities:
%		\begin{enumerate}
\item $\bar{C}(T_k)  = \emptyset$, so that $T_k$ is a star graph with a central node and three neighboring leaves. It holds that $c(T_k)=2$. Removing one leaf $v$, $T_{k}^v$ is a path graph of length 2, therefore $c(T_{k}^v)=2=c(T_{k})$ (see Figure \ref{fig:inttarget4}).
\item $\bar{C}(T_k) \not = \emptyset$: then,   it holds that $c_1(T_{k})= c_2(T_{k})$ and, consequently, $c(T_k)=c_2(T_k)+2$. Moreover, there exists at least one corridor in  $\bar{C}(T_k)$ such that one of its endpoints is a junction connected to exactly two leaves, since $L_2(T_k)=L_4(T_k)=\emptyset$. We choose $v$ between one of the two leaves (see Figure \ref{fig:inttarget3}). In this case, $c_1(T_{k}^v) \leq c_2(T_{k})+1$ and $c_2(T_{k}^v) \leq c_2(T_{k})$, therefore 
			\[c(T_{k}^v)=\max \{c_1(T_{k}^v)+1, c_2(T_{k}^v)+2\} \leq \]
			\[ \leq \max \{(c_2(T_{k})+1)+1,c_2(T_{k})+2\} = c(T_k). \]
				%\item $T_k$ is a "star": it has a central node with three neighboring leaves. It holds that $c(T_k)=2$. Removing one leaf, $T_{k}^v$ is a path graph of length 2, therefore $c(T_{k}^v)=2=c(T_{k})$.
		\end{enumerate} 

	\end{itemize}
	
\end{proof}

We propose the following \textit{Leaves procedure for PMT}, which breaks down the PMT problem into an unlabeled problem and a series of motion planning problems, and find plan $f$ for any PMT instance.\\

%\vspace{1.5cm}

\noindent \textit{Leaves procedure for} PMT
\begin{enumerate}
	\item Let $V_1$ be the set of intermediate targets found with the previous procedure. Find a plan $g$ which solves the unlabeled PMT problem from the final configuration $\A^t$ to $V_1$, i.e., such that
	\[ V_1 = \rho(\A^t,g)(P),\]
	
	and let $\A^{\bar{t}}:= \rho(\A^t,g)$ be the \textit{intermediate configuration}. Note that for all $k\in \{1,\dots,  |P|\}$, $g$ would move a pebble from  $t_{i_k}$ to the intermediate target $\bar{t}_{k}$.
	
%	\item \bndel{Sort the nodes of $V_1$ from closest to farthest to the leaves, i.e., the ordered sequence of the intermediate targets $(\bar{t}_{i_1},\dots, \bar{t}_{i_{ |P|}})$ is such that }
	
%\bndel{	$ d(\bar{t}_{i_k},L(T)) \leq  d(\bar{t}_{i_{k+1}},L(T)) \quad \forall k=1,\dots,  |P|;$    }
	
	\item Set $k=1$ and perform the following procedure:
	\begin{enumerate}
		\item if $k>|P|$, stop;
		\item using CATERPILLAR algorithm, solve the motion planning for pebble $p_{i_k}$ from $\A^s(p_{i_k})$ to %\bndel{$\bar{t}_{i_k}$}
		   $\bar{t}_{k}$, i.e., find a plan $f_{k}$, over the tree $T_k$ obtained from $T$ by removing nodes %\bndel{$\bar{t}_{i_1},\dots,\bar{t}_{i_{k-1}}$}
		  $\bar{t}_{1},\dots,\bar{t}_{{k-1}}$, such that %\bndel{$\bar{t}_{i_k}$}
		     $\bar{t}_{k}$ $= \rho(\A^s,f_{k})(p_{i_k})$;
		
		\item update $\A^s = \rho(\A^s,f_{k})$;
		\item set $k = k+1$ and go back to Step $a)$.
		
	\end{enumerate}
	 
	\item Apply $g^{-1}$, the inverse plan of $g$, which,  for all $k\in \{1,\dots,  |P|\}$, moves pebble $p_{i_k}$ from  %\bndel{$\bar{t}_{i_k}$} 
	 $\bar{t}_{k}$ to $t_{i_k}$.
	\end{enumerate}

Therefore, plan $f$ which solves a given PMT instance is
\begin{equation}
	\label{solPMT}
f = f_{1} f_{2} \cdots f_{{ |P|}} \, g^{-1}.
\end{equation}

The complexity of the proposed procedure is stated in the following theorem.
\begin{theorem}
\label{th:pmt_comp}
The length complexity of the proposed procedure for the solution of the PMT problem is $O(|P|n c+n^2)$.
\end{theorem}
\begin{proof}
Note that $f_{k}$ requires at most $O(n c)$ moves for all $k=1, \cdots, |P|$. Indeed, each $f_{k}$ is the solution of a motion planning problem, which, in view of Propositions \ref{prop:procA} and \ref{prop:procB}, is solved in $O(nc)$ moves. Moreover, $g^{-1}$ requires at most $n^2$ moves. Indeed, $g$ is the solution of an unlabeled PMT, which requires $O(n^2)$ moves, as seen in Section \ref{sec:upmt}. Since $g^{-1}$ is obtained from $g$ by reversing its moves, it has the same length (see Observation \ref{reverse}). Then, the total number of moves is 
 \[  O(|P| n c) + O(n^2).\]
\end{proof}
Figure \ref{fig:pmt} provides an example of application of the  \textit{Leaves procedure} for a PMT instance with three pebbles.
% \vspace{5cm}
 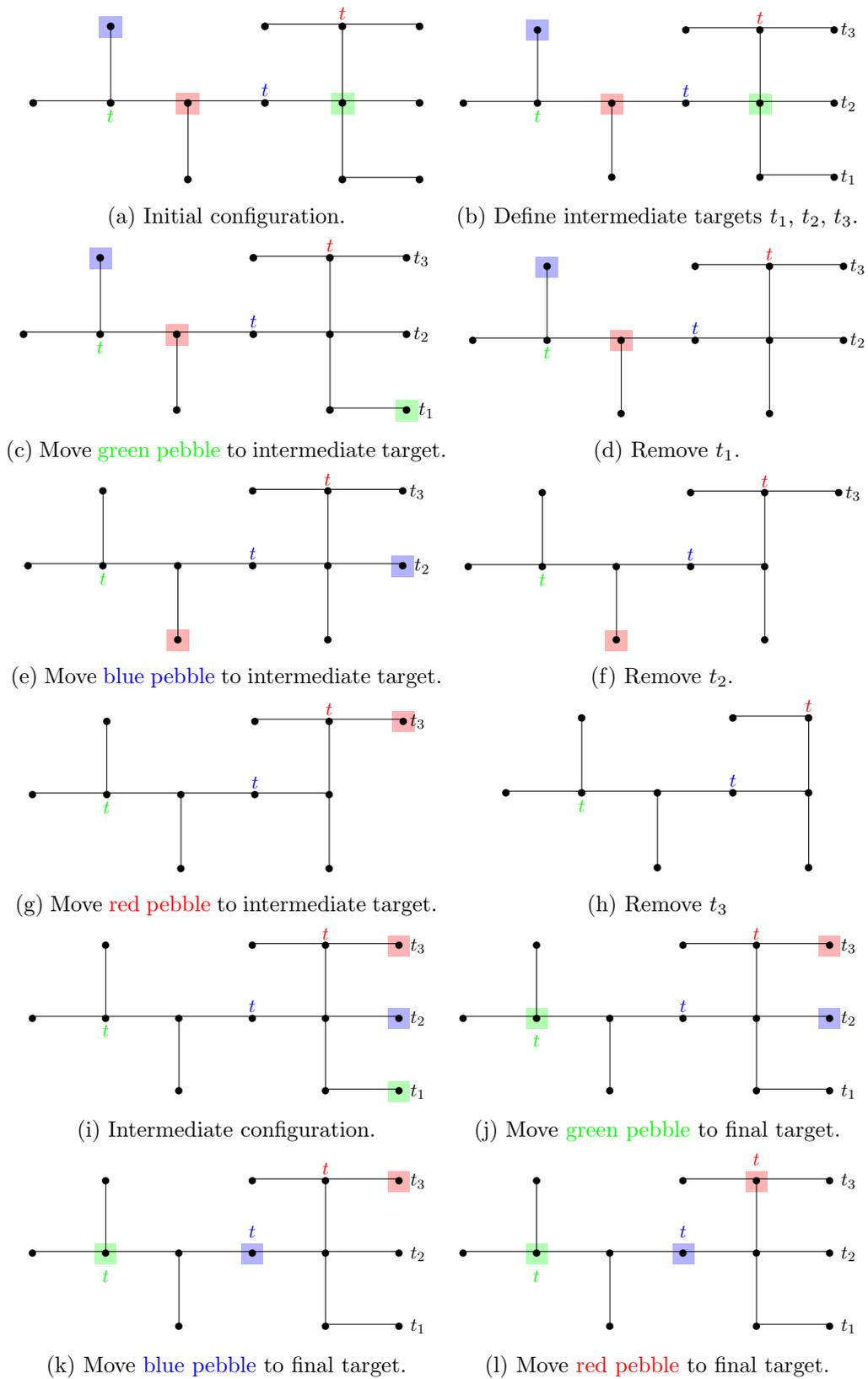
\begin{figure}[htp!]
	\centering
	\begin{subfigure}[t]{0.45\columnwidth}
		\centering
		\resizebox{\columnwidth}{!}{
			\begin{tikzpicture}
				\node (A) at (0,0) {\textbullet};
				\node (B) at (1.5,0) {\textbullet};
				\node[below] at (B) {{\color{green}$t$}};
				
				\node[style={rectangle,fill=blue!30}] (J) at (1.5,1.5) {\textbullet};
				\node[style={rectangle,fill=red!30}] (C) at (3,0) {\textbullet};
				%	\node[below] at (A) {$s$};
				%	\node[above] at (A) {$f_0$};
				%	\node[below] at (B) {$i_0$};
				%	\node[above] at (C) {$i_1$};
				\node (K) at (3,-1.5) {\textbullet};
				
				\draw  (1.5,0) -- (1.5,1.5);
				\draw  (3,0) -- (3,-1.5);
				%\node[left] at (K) {$f_1$};
				\node(D)at (4.5,0) {\textbullet};
				\node[above] at (D) {{\color{blue}$t$}};
				\node[style={rectangle,fill=green!30}]  (E) at (6,0) {\textbullet};
				
				\node (F) at (7.5,0) {\textbullet} ;
				%	\node[right] at (F) {$t_2$};
				
				\node   (L) at (7.5,-1.5) {\textbullet};
				%	\node[right] at (L) {$t_1$};
				%	\node[left] at (L) {$f_2$};
				%\draw (7.5,0) -- (7.5,-1.5);
				%	\node[above] at (F) {$j_1=i_2$};
				
				\node (I) at (6,1.5) {\textbullet};
				
				\node[above] at (I) {{\color{red}$t$}};
				\node  (N) at (4.5,1.5) {\textbullet};
				\node (O) at (7.5,1.5) {\textbullet};
				%	\node[right] at (O) {$t_3$};
				\draw  (0,0.05) -- (7.5,0.05);
				\draw  (4.5,1.55) -- (7.5,1.55);
				\draw  (6,0) -- (6,1.5);
				
				\node (P) at (6,-1.5) {\textbullet};
				
				\draw (6,-1.45) -- (7.5,-1.45); 	
				\draw (6,0) -- (6,-1.5);

			\end{tikzpicture}
		}
		\caption{Initial configuration. }
		\label{fig:pmt}
	\end{subfigure}
	\begin{subfigure}[t]{0.45\columnwidth}
		\centering
		\resizebox{\columnwidth}{!}{
			\begin{tikzpicture}
				\node (A) at (0,0) {\textbullet};
				\node (B) at (1.5,0) {\textbullet};
				\node[below] at (B) {{\color{green}$t$}};
				
				\node[style={rectangle,fill=blue!30}] (J) at (1.5,1.5) {\textbullet};
				\node[style={rectangle,fill=red!30}] (C) at (3,0) {\textbullet};
				%	\node[below] at (A) {$s$};
				%	\node[above] at (A) {$f_0$};
				%	\node[below] at (B) {$i_0$};
				%	\node[above] at (C) {$i_1$};
				\node (K) at (3,-1.5) {\textbullet};
				
				\draw  (1.5,0) -- (1.5,1.5);
				\draw  (3,0) -- (3,-1.5);
				%\node[left] at (K) {$f_1$};
				\node(D)at (4.5,0) {\textbullet};
				\node[above] at (D) {{\color{blue}$t$}};
				\node[style={rectangle,fill=green!30}]  (E) at (6,0) {\textbullet};
				
				\node (F) at (7.5,0) {\textbullet} ;
				\node[right] at (F) {$t_2$};
				
				\node   (L) at (7.5,-1.5) {\textbullet};
				\node[right] at (L) {$t_1$};
				%	\node[left] at (L) {$f_2$};
				%\draw (7.5,0) -- (7.5,-1.5);
				%	\node[above] at (F) {$j_1=i_2$};
				
				\node (I) at (6,1.5) {\textbullet};
				
				\node[above] at (I) {{\color{red}$t$}};
				\node  (N) at (4.5,1.5) {\textbullet};
				\node (O) at (7.5,1.5) {\textbullet};
				\node[right] at (O) {$t_3$};
				\draw  (0,0.05) -- (7.5,0.05);
				\draw  (4.5,1.55) -- (7.5,1.55);
				\draw  (6,0) -- (6,1.5);
				
				\node (P) at (6,-1.5) {\textbullet};
				
				\draw (6,-1.45) -- (7.5,-1.45); 	
				\draw (6,0) -- (6,-1.5);

			\end{tikzpicture}
		}
		\caption{Define intermediate targets $t_1$, $t_2,$ $t_3$. }
		\label{pmt1}
	\end{subfigure}
	\begin{subfigure}[t]{0.47\columnwidth}
		\centering
		\resizebox{\columnwidth}{!}{
			\begin{tikzpicture}
				\node (A) at (0,0) {\textbullet};
				\node (B) at (1.5,0) {\textbullet};
				\node[below] at (B) {{\color{green}$t$}};
				
				\node[style={rectangle,fill=blue!30}] (J) at (1.5,1.5) {\textbullet};
				\node[style={rectangle,fill=red!30}] (C) at (3,0) {\textbullet};
				%	\node[below] at (A) {$s$};
				%	\node[above] at (A) {$f_0$};
				%	\node[below] at (B) {$i_0$};
				%	\node[above] at (C) {$i_1$};
				\node (K) at (3,-1.5) {\textbullet};
				
				\draw  (1.5,0) -- (1.5,1.5);
				\draw  (3,0) -- (3,-1.5);
				%\node[left] at (K) {$f_1$};
				\node(D)at (4.5,0) {\textbullet};
				\node[above] at (D) {{\color{blue}$t$}};
				\node  (E) at (6,0) {\textbullet};
				
				\node (F) at (7.5,0) {\textbullet} ;
				\node[right] at (F) {$t_2$};
				
				\node[style={rectangle,fill=green!30}]   (L) at (7.5,-1.5) {\textbullet};
				\node[right] at (7.6,-1.5) {$t_1$};
				%	\node[left] at (L) {$f_2$};
				%\draw (7.5,0) -- (7.5,-1.5);
				%	\node[above] at (F) {$j_1=i_2$};
				
				\node (I) at (6,1.5) {\textbullet};
				
				\node[above] at (I) {{\color{red}$t$}};
				\node  (N) at (4.5,1.5) {\textbullet};
				\node (O) at (7.5,1.5) {\textbullet};
				\node[right] at (O) {$t_3$};
				\draw  (0,0.05) -- (7.5,0.05);
				\draw  (4.5,1.55) -- (7.5,1.55);
				\draw  (6,0) -- (6,1.5);
				
				\node (P) at (6,-1.5) {\textbullet};
				
				\draw (6,-1.45) -- (7.5,-1.45); 	
				\draw (6,0) -- (6,-1.5);

			\end{tikzpicture}
		}
		\caption{Move {\color{green} green pebble} to intermediate target. }
		\label{pmt2}
	\end{subfigure}
	\begin{subfigure}[t]{0.45\columnwidth}
		\centering
		\resizebox{\columnwidth}{!}{
			\begin{tikzpicture}
				\node (A) at (0,0) {\textbullet};
				\node (B) at (1.5,0) {\textbullet};
				\node[below] at (B) {{\color{green}$t$}};
				
				\node[style={rectangle,fill=blue!30}] (J) at (1.5,1.5) {\textbullet};
				\node[style={rectangle,fill=red!30}] (C) at (3,0) {\textbullet};
				%	\node[below] at (A) {$s$};
				%	\node[above] at (A) {$f_0$};
				%	\node[below] at (B) {$i_0$};
				%	\node[above] at (C) {$i_1$};
				\node (K) at (3,-1.5) {\textbullet};
				
				\draw  (1.5,0) -- (1.5,1.5);
				\draw  (3,0) -- (3,-1.5);
				%\node[left] at (K) {$f_1$};
				\node(D)at (4.5,0) {\textbullet};
				\node[above] at (D) {{\color{blue}$t$}};
				\node  (E) at (6,0) {\textbullet};
				
				\node (F) at (7.5,0) {\textbullet} ;
				\node[right] at (F) {$t_2$};
				
				%	\node[style={rectangle,fill=green!30}]   (L) at (7.5,-1.5) {\textbullet};
				%	\node[right] at (7.6,-1.5) {$t_1$};
				%	\node[left] at (L) {$f_2$};
				%\draw (7.5,0) -- (7.5,-1.5);
				%	\node[above] at (F) {$j_1=i_2$};
				
				\node (I) at (6,1.5) {\textbullet};
				
				\node[above] at (I) {{\color{red}$t$}};
				\node  (N) at (4.5,1.5) {\textbullet};
				\node (O) at (7.5,1.5) {\textbullet};
				\node[right] at (O) {$t_3$};
				\draw  (0,0.05) -- (7.5,0.05);
				\draw  (4.5,1.55) -- (7.5,1.55);
				\draw  (6,0) -- (6,1.5);
				
				\node (P) at (6,-1.5) {\textbullet};
				
				%	\draw (6,-1.45) -- (7.5,-1.45); 	
				\draw (6,0) -- (6,-1.5);

			\end{tikzpicture}
		}
		\caption{Remove $t_1$.}
		
		\label{pmt3}
	\end{subfigure}
	\begin{subfigure}[t]{0.46\columnwidth}
		\centering
		\resizebox{\columnwidth}{!}{
			\begin{tikzpicture}
				\node (A) at (0,0) {\textbullet};
				\node (B) at (1.5,0) {\textbullet};
				\node[below] at (B) {{\color{green}$t$}};
				
				\node (J) at (1.5,1.5) {\textbullet};
				\node (C) at (3,0) {\textbullet};
				%	\node[below] at (A) {$s$};
				%	\node[above] at (A) {$f_0$};
				%	\node[below] at (B) {$i_0$};
				%	\node[above] at (C) {$i_1$};
				\node[style={rectangle,fill=red!30}] (K) at (3,-1.5) {\textbullet};
				
				\draw  (1.5,0) -- (1.5,1.5);
				\draw  (3,0) -- (3,-1.5);
				%\node[left] at (K) {$f_1$};
				\node(D)at (4.5,0) {\textbullet};
				\node[above] at (D) {{\color{blue}$t$}};
				\node  (E) at (6,0) {\textbullet};
				
				\node[style={rectangle,fill=blue!30}] (F) at (7.5,0) {\textbullet} ;
				\node[right] at (7.6,0) {$t_2$};
				
				%	\node[style={rectangle,fill=green!30}]   (L) at (7.5,-1.5) {\textbullet};
				%	\node[right] at (7.6,-1.5) {$t_1$};
				%	\node[left] at (L) {$f_2$};
				%\draw (7.5,0) -- (7.5,-1.5);
				%	\node[above] at (F) {$j_1=i_2$};
				
				\node (I) at (6,1.5) {\textbullet};
				
				\node[above] at (I) {{\color{red}$t$}};
				\node  (N) at (4.5,1.5) {\textbullet};
				\node (O) at (7.5,1.5) {\textbullet};
				\node[right] at (O) {$t_3$};
				\draw  (0,0.05) -- (7.5,0.05);
				\draw  (4.5,1.55) -- (7.5,1.55);
				\draw  (6,0) -- (6,1.5);
				
				\node (P) at (6,-1.5) {\textbullet};
				
				%	\draw (6,-1.45) -- (7.5,-1.45); 	
				\draw (6,0) -- (6,-1.5);

			\end{tikzpicture}
		}
		\caption{Move {\color{blue} blue pebble} to intermediate target. }
		\label{pmt4}
	\end{subfigure}
	\begin{subfigure}[t]{0.45\columnwidth}
		\centering
		\resizebox{\columnwidth}{!}{
			\begin{tikzpicture}
				\node (A) at (0,0) {\textbullet};
				\node (B) at (1.5,0) {\textbullet};
				\node[below] at (B) {{\color{green}$t$}};
				
				\node (J) at (1.5,1.5) {\textbullet};
				\node (C) at (3,0) {\textbullet};
				%	\node[below] at (A) {$s$};
				%	\node[above] at (A) {$f_0$};
				%	\node[below] at (B) {$i_0$};
				%	\node[above] at (C) {$i_1$};
				\node[style={rectangle,fill=red!30}] (K) at (3,-1.5) {\textbullet};
				
				\draw  (1.5,0) -- (1.5,1.5);
				\draw  (3,0) -- (3,-1.5);
				%\node[left] at (K) {$f_1$};
				\node(D)at (4.5,0) {\textbullet};
				\node[above] at (D) {{\color{blue}$t$}};
				\node  (E) at (6,0) {\textbullet};
				
				%	\node[style={rectangle,fill=blue!30}] (F) at (7.5,0) {\textbullet} ;
				%	\node[right] at (F) {$t_2$};
				
				%	\node[style={rectangle,fill=green!30}]   (L) at (7.5,-1.5) {\textbullet};
				%	\node[right] at (7.6,-1.5) {$t_1$};
				%	\node[left] at (L) {$f_2$};
				%\draw (7.5,0) -- (7.5,-1.5);
				%	\node[above] at (F) {$j_1=i_2$};
				
				\node (I) at (6,1.5) {\textbullet};
				
				\node[above] at (I) {{\color{red}$t$}};
				\node  (N) at (4.5,1.5) {\textbullet};
				\node (O) at (7.5,1.5) {\textbullet};
				\node[right] at (O) {$t_3$};
				\draw  (0,0.05) -- (6,0.05);
				\draw  (4.5,1.55) -- (7.5,1.55);
				\draw  (6,0) -- (6,1.5);
				
				\node (P) at (6,-1.5) {\textbullet};
				
				%	\draw (6,-1.45) -- (7.5,-1.45); 	
				\draw (6,0) -- (6,-1.5);

			\end{tikzpicture}
		}
		\caption{Remove $t_2$.}
		\label{pmt5}
	\end{subfigure}
	\begin{subfigure}[t]{0.45\columnwidth}
		\centering
		\resizebox{\columnwidth}{!}{
			\begin{tikzpicture}
				\node (A) at (0,0) {\textbullet};
				\node (B) at (1.5,0) {\textbullet};
				\node[below] at (B) {{\color{green}$t$}};
				
				\node (J) at (1.5,1.5) {\textbullet};
				\node (C) at (3,0) {\textbullet};
				%	\node[below] at (A) {$s$};
				%	\node[above] at (A) {$f_0$};
				%	\node[below] at (B) {$i_0$};
				%	\node[above] at (C) {$i_1$};
				\node (K) at (3,-1.5) {\textbullet};
				
				\draw  (1.5,0) -- (1.5,1.5);
				\draw  (3,0) -- (3,-1.5);
				%\node[left] at (K) {$f_1$};
				\node(D)at (4.5,0) {\textbullet};
				\node[above] at (D) {{\color{blue}$t$}};
				\node  (E) at (6,0) {\textbullet};
				
				%	\node[style={rectangle,fill=blue!30}] (F) at (7.5,0) {\textbullet} ;
				%	\node[right] at (F) {$t_2$};
				
				%	\node[style={rectangle,fill=green!30}]   (L) at (7.5,-1.5) {\textbullet};
				%	\node[right] at (7.6,-1.5) {$t_1$};
				%	\node[left] at (L) {$f_2$};
				%\draw (7.5,0) -- (7.5,-1.5);
				%	\node[above] at (F) {$j_1=i_2$};
				
				\node (I) at (6,1.5) {\textbullet};
				
				\node[above] at (I) {{\color{red}$t$}};
				\node  (N) at (4.5,1.5) {\textbullet};
				\node[style={rectangle,fill=red!30}] (O) at (7.5,1.5) {\textbullet};
				\node[right] at (O) {$t_3$};
				\draw  (0,0.05) -- (6,0.05);
				\draw  (4.5,1.55) -- (7.5,1.55);
				\draw  (6,0) -- (6,1.5);
				
				\node (P) at (6,-1.5) {\textbullet};
				
				%	\draw (6,-1.45) -- (7.5,-1.45); 	
				\draw (6,0) -- (6,-1.5);

			\end{tikzpicture}
		}
		\caption{Move {\color{red} red pebble} to intermediate target. }
		\label{pmt6}
	\end{subfigure}
	\begin{subfigure}[t]{0.45\columnwidth}
		\centering
		\resizebox{0.8\columnwidth}{!}{
			\begin{tikzpicture}
				\node (A) at (0,0) {\textbullet};
				\node (B) at (1.5,0) {\textbullet};
				\node[below] at (B) {{\color{green}$t$}};
				
				\node (J) at (1.5,1.5) {\textbullet};
				\node (C) at (3,0) {\textbullet};
				%	\node[below] at (A) {$s$};
				%	\node[above] at (A) {$f_0$};
				%	\node[below] at (B) {$i_0$};
				%	\node[above] at (C) {$i_1$};
				\node (K) at (3,-1.5) {\textbullet};
				
				\draw  (1.5,0) -- (1.5,1.5);
				\draw  (3,0) -- (3,-1.5);
				%\node[left] at (K) {$f_1$};
				\node(D)at (4.5,0) {\textbullet};
				\node[above] at (D) {{\color{blue}$t$}};
				\node  (E) at (6,0) {\textbullet};
				
				%	\node[style={rectangle,fill=blue!30}] (F) at (7.5,0) {\textbullet} ;
				%	\node[right] at (F) {$t_2$};
				
				%	\node[style={rectangle,fill=green!30}]   (L) at (7.5,-1.5) {\textbullet};
				%	\node[right] at (7.6,-1.5) {$t_1$};
				%	\node[left] at (L) {$f_2$};
				%\draw (7.5,0) -- (7.5,-1.5);
				%	\node[above] at (F) {$j_1=i_2$};
				
				\node (I) at (6,1.5) {\textbullet};
				
				\node[above] at (I) {{\color{red}$t$}};
				\node  (N) at (4.5,1.5) {\textbullet};
				%	\node[style={rectangle,fill=red!30}] (O) at (7.5,1.5) {\textbullet};
				%	\node[right] at (O) {$t_3$};
				\draw  (0,0.05) -- (6,0.05);
				\draw  (4.5,1.55) -- (6,1.55);
				\draw  (6,0) -- (6,1.5);
				
				\node (P) at (6,-1.5) {\textbullet};
				
				%	\draw (6,-1.45) -- (7.5,-1.45); 	
				\draw (6,0) -- (6,-1.5);

			\end{tikzpicture}
		}
		\caption{Remove $t_3$ }
		\label{pmt7}
	\end{subfigure}
	\begin{subfigure}[t]{0.45\columnwidth}
		\centering
		\resizebox{\columnwidth}{!}{
			\begin{tikzpicture}
				\node (A) at (0,0) {\textbullet};
				\node (B) at (1.5,0) {\textbullet};
				\node[below] at (B) {{\color{green}$t$}};
				
				\node (J) at (1.5,1.5) {\textbullet};
				\node (C) at (3,0) {\textbullet};
				%	\node[below] at (A) {$s$};
				%	\node[above] at (A) {$f_0$};
				%	\node[below] at (B) {$i_0$};
				%	\node[above] at (C) {$i_1$};
				\node (K) at (3,-1.5) {\textbullet};
				
				\draw  (1.5,0) -- (1.5,1.5);
				\draw  (3,0) -- (3,-1.5);
				%\node[left] at (K) {$f_1$};
				\node(D)at (4.5,0) {\textbullet};
				\node[above] at (D) {{\color{blue}$t$}};
				\node  (E) at (6,0) {\textbullet};
				
				\node[style={rectangle,fill=blue!30}] (F) at (7.5,0) {\textbullet} ;
				\node[right] at (7.6,0) {$t_2$};
				
				\node[style={rectangle,fill=green!30}]   (L) at (7.5,-1.5) {\textbullet};
				\node[right] at (7.6,-1.5) {$t_1$};
				%	\node[left] at (L) {$f_2$};
				%\draw (7.5,0) -- (7.5,-1.5);
				%	\node[above] at (F) {$j_1=i_2$};
				
				\node (I) at (6,1.5) {\textbullet};
				
				\node[above] at (I) {{\color{red}$t$}};
				\node  (N) at (4.5,1.5) {\textbullet};
				\node[style={rectangle,fill=red!30}] (O) at (7.5,1.5) {\textbullet};
				\node[right] at (7.6,1.5) {$t_3$};
				\draw  (0,0.05) -- (7.5,0.05);
				\draw  (4.5,1.55) -- (7.5,1.55);
				\draw  (6,0) -- (6,1.5);
				
				\node (P) at (6,-1.5) {\textbullet};
				
				\draw (6,-1.45) -- (7.5,-1.45); 	
				\draw (6,0) -- (6,-1.5);

			\end{tikzpicture}
		}
		\caption{Intermediate configuration. }
		\label{pmt8}
	\end{subfigure}
	\begin{subfigure}[t]{0.45\columnwidth}
		\centering
		\resizebox{\columnwidth}{!}{
			\begin{tikzpicture}
				\node (A) at (0,0) {\textbullet};
				\node[style={rectangle,fill=green!30}]  (B) at (1.5,0) {\textbullet};
				\node[below] at (1.5,-0.2) {{\color{green}$t$}};
				
				\node (J) at (1.5,1.5) {\textbullet};
				\node (C) at (3,0) {\textbullet};
				%	\node[below] at (A) {$s$};
				%	\node[above] at (A) {$f_0$};
				%	\node[below] at (B) {$i_0$};
				%	\node[above] at (C) {$i_1$};
				\node (K) at (3,-1.5) {\textbullet};
				
				\draw  (1.5,0) -- (1.5,1.5);
				\draw  (3,0) -- (3,-1.5);
				%\node[left] at (K) {$f_1$};
				\node(D)at (4.5,0) {\textbullet};
				\node[above] at (D) {{\color{blue}$t$}};
				\node  (E) at (6,0) {\textbullet};
				
				\node[style={rectangle,fill=blue!30}] (F) at (7.5,0) {\textbullet} ;
				\node[right] at (7.6,0) {$t_2$};
				
				\node  (L) at (7.5,-1.5) {\textbullet};
				\node[right] at (7.6,-1.5) {$t_1$};
				%	\node[left] at (L) {$f_2$};
				%\draw (7.5,0) -- (7.5,-1.5);
				%	\node[above] at (F) {$j_1=i_2$};
				
				\node (I) at (6,1.5) {\textbullet};
				
				\node[above] at (I) {{\color{red}$t$}};
				\node  (N) at (4.5,1.5) {\textbullet};
				\node[style={rectangle,fill=red!30}] (O) at (7.5,1.5) {\textbullet};
				\node[right] at (7.6,1.5) {$t_3$};
				\draw  (0,0.05) -- (7.5,0.05);
				\draw  (4.5,1.55) -- (7.5,1.55);
				\draw  (6,0) -- (6,1.5);
				
				\node (P) at (6,-1.5) {\textbullet};
				
				\draw (6,-1.45) -- (7.5,-1.45); 	
				\draw (6,0) -- (6,-1.5);

			\end{tikzpicture}
		}
		\caption{Move {\color{green} green pebble} to final target. }
		\label{pmt9}
	\end{subfigure}
	\begin{subfigure}[t]{0.45\columnwidth}
		\centering
		\resizebox{\columnwidth}{!}{
			\begin{tikzpicture}
				\node (A) at (0,0) {\textbullet};
				\node[style={rectangle,fill=green!30}]  (B) at (1.5,0) {\textbullet};
				\node[below] at (1.5,-0.2) {{\color{green}$t$}};
				
				\node (J) at (1.5,1.5) {\textbullet};
				\node (C) at (3,0) {\textbullet};
				%	\node[below] at (A) {$s$};
				%	\node[above] at (A) {$f_0$};
				%	\node[below] at (B) {$i_0$};
				%	\node[above] at (C) {$i_1$};
				\node (K) at (3,-1.5) {\textbullet};
				
				\draw  (1.5,0) -- (1.5,1.5);
				\draw  (3,0) -- (3,-1.5);
				%\node[left] at (K) {$f_1$};
				\node[style={rectangle,fill=blue!30}] (D) at (4.5,0) {\textbullet};
				\node[above] at (4.5,0.2) {{\color{blue}$t$}};
				\node  (E) at (6,0) {\textbullet};
				
				\node (F) at (7.5,0) {\textbullet} ;
				\node[right] at (7.6,0) {$t_2$};
				
				\node  (L) at (7.5,-1.5) {\textbullet};
				\node[right] at (7.6,-1.5) {$t_1$};
				%	\node[left] at (L) {$f_2$};
				%\draw (7.5,0) -- (7.5,-1.5);
				%	\node[above] at (F) {$j_1=i_2$};
				
				\node (I) at (6,1.5) {\textbullet};
				
				\node[above] at (I) {{\color{red}$t$}};
				\node  (N) at (4.5,1.5) {\textbullet};
				\node[style={rectangle,fill=red!30}] (O) at (7.5,1.5) {\textbullet};
				\node[right] at (7.6,1.5) {$t_3$};
				\draw  (0,0.05) -- (7.5,0.05);
				\draw  (4.5,1.55) -- (7.5,1.55);
				\draw  (6,0) -- (6,1.5);
				
				\node (P) at (6,-1.5) {\textbullet};
				
				\draw (6,-1.45) -- (7.5,-1.45); 	
				\draw (6,0) -- (6,-1.5);

			\end{tikzpicture}
		}
		\caption{Move {\color{blue} blue pebble} to final target. }
		\label{pmt10}
	\end{subfigure}
	\begin{subfigure}[t]{0.45\columnwidth}
		\centering
		\resizebox{\columnwidth}{!}{
			\begin{tikzpicture}
				\node (A) at (0,0) {\textbullet};
				\node[style={rectangle,fill=green!30}]  (B) at (1.5,0) {\textbullet};
				\node[below] at (1.5,-0.2) {{\color{green}$t$}};
				
				\node (J) at (1.5,1.5) {\textbullet};
				\node (C) at (3,0) {\textbullet};
				%	\node[below] at (A) {$s$};
				%	\node[above] at (A) {$f_0$};
				%	\node[below] at (B) {$i_0$};
				%	\node[above] at (C) {$i_1$};
				\node (K) at (3,-1.5) {\textbullet};
				
				\draw  (1.5,0) -- (1.5,1.5);
				\draw  (3,0) -- (3,-1.5);
				%\node[left] at (K) {$f_1$};
				\node[style={rectangle,fill=blue!30}] (D) at (4.5,0) {\textbullet};
				\node[above] at (4.5,0.2) {{\color{blue}$t$}};
				\node  (E) at (6,0) {\textbullet};
				
				\node (F) at (7.5,0) {\textbullet} ;
				\node[right] at (7.6,0) {$t_2$};
				
				\node  (L) at (7.5,-1.5) {\textbullet};
				\node[right] at (7.6,-1.5) {$t_1$};
				%	\node[left] at (L) {$f_2$};
				%\draw (7.5,0) -- (7.5,-1.5);
				%	\node[above] at (F) {$j_1=i_2$};
				
				\node[style={rectangle,fill=red!30}] (I) at (6,1.5) {\textbullet};
				
				\node[above] at (6,1.7) {{\color{red}$t$}};
				\node  (N) at (4.5,1.5) {\textbullet};
				\node (O) at (7.5,1.5) {\textbullet};
				\node[right] at (7.6,1.5) {$t_3$};
				\draw  (0,0.05) -- (7.5,0.05);
				\draw  (4.5,1.55) -- (7.5,1.55);
				\draw  (6,0) -- (6,1.5);
				
				\node (P) at (6,-1.5) {\textbullet};
				
				\draw (6,-1.45) -- (7.5,-1.45); 	
				\draw (6,0) -- (6,-1.5);

			\end{tikzpicture}
		}
		\caption{Move {\color{red} red pebble} to final target. }
		\label{pmt11}
	\end{subfigure}
	\caption{Example of application of the \textit{Leaves procedure}.}.
\end{figure}
$\ $\newline\newline\noindent
An interesting property of the \textit{Leaves procedure for} PMT concerns the number of times each vertex is traversed by pebbles. As we further discuss in Section \ref{tspmt}, the pebble motion problem on general graphs can be solved after converting it on a variant of PMT over trees. The bound on the number of times each vertex is crossed by the pebbles in the PMT problem over trees provided by the following proposition, allows deriving  a complexity result also for the pebble motion problem over general graphs. Such complexity result will be the topic of a forthcoming paper.
\begin{prop}
 	In any solution provided by the proposed procedure, each vertex is crossed $O(|P|c)$ times by the pebbles.
 \end{prop}  
\begin{proof}

	Let us count how many times each vertex is crossed in the solutions of each problem described in the paper:
	\begin{enumerate}
	
			\item \textit{Basic plans}. 
			\begin{itemize}
				\item  \textsc{Bring hole from $w$ to $v$} ($\alpha_{vw}$): each pebble along path $\pi_{vw}$ moves forward one position, therefore each vertex is crossed at most once.
				\item \textsc{Move Pebble from $v$ to $w$} ($\beta_{vw}$): one pebble moves on the path $\pi_{vw}$, therefore each vertex is crossed at most once.
			\end{itemize}

		\item \textit{Unlabeled} PMT \textit{problem}. In the solution of the \textit{unlabeled} problem proposed in Section \ref{sec:upmt}, each  vertex is crossed at most once by each pebble for a total of $O(|P|)$ times in the overall procedure.

		\item \textit{Gather c holes}. In each iteration of Step 2 of the procedure described in Section \ref{sec_gather}, each vertex is crossed at most once because it performs $\alpha_{uv}$ or $\beta_{uw} \alpha_{wv}$. Since Step 2 is executed at most $c$ times, each vertex is crossed $O(c)$ times.
		\item CATERPILLAR algorithm. Each vertex is crossed $O(c)$ times, indeed:
		
		\begin{enumerate}

\item \textit{Procedure A}. In Step 1 we solve a \textit{gather hole problem}, therefore by point (2) each vertex is crossed $O(c)$ times. In Step 3 each vertex that belongs to a caterpillar set $S_k$ is traversed once time by the pebble and once time by at most $c$ obstacles that arrived from $S_{k+1}$ and then moved to $S_{k-1}$. Therefore, in \textit{Procedure A} each vertex is crossed $O(c)$ times.
			\item \textit{Procedure B}. In Step 2 we solve a  \textit{gather hole problem} with $q_j$ holes, therefore each vertex is crossed $O(q_j)$ times. Since $\sum_{j=1}^k q_j = q \leq c$, each vertex is crossed at most $O(c)$ times.
	
		\end{enumerate}
		
		\item \textit{Leaves Procedure.} A solution of PMT given by this procedure is $f = f_{1} f_{2} \cdots f_{{ |P|}} \, g^{-1}$ (see (\ref{solPMT})). For each $k \in \{1, \dots, |P|\}$, $f_k$ is the solution of a motion planning problem provided by the CATERPILLAR algorithm, therefore each vertex is crossed $O(c)$ times. Moreover, $g^{-1}$ is the inverse of the solution of an unlabeled problem: therefore each vertex is crossed $O(|P|)$ times. We can conclude that in the solution plan of PMT  each vertex is crossed
		\[|P| O(c) + O(|P|) = O(|P| c)\]
		times.
	
\end{enumerate}

	\end{proof}

\section{PMT with Trans-shipment vertices}
\label{tspmt}
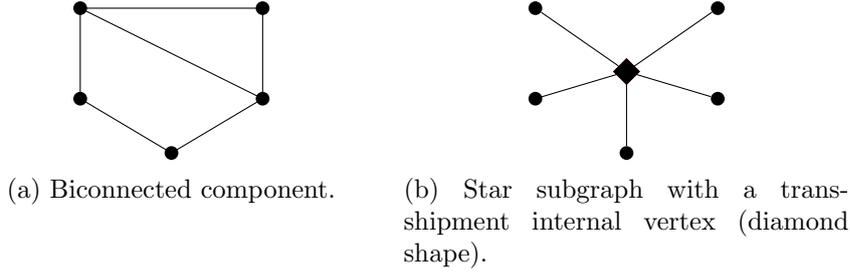
\begin{figure}[h!]
	\centering
	%	\resizebox{\columnwidth}{!}{

		\begin{subfigure}[t]{0.4\textwidth}
			\centering
			\begin{tikzpicture}
				[scale=1.2,auto=left]
			    \node[style={circle,fill=black,scale=0.5}] (A) at (0,0) {};
			    
			    \node[style={circle,fill=black,scale=0.5}]  (B) at (1,-0.6)  {};
			    
			    \node[style={circle,fill=black,scale=0.5}]  (C) at (2,0) {};
			    
			    \node[style={circle,fill=black,scale=0.5}]  (D) at (2,1) {};
			    
			    \node[style={circle,fill=black,scale=0.5}]  (E) at (0,1) {};

				\draw  (0,0) -- (1,-0.6);
				\draw  (1,-0.6) -- (2,0);
			    \draw (2,0) -- (2,1);
			    
			   \draw (2,1) -- (0,1);
			   \draw (0,1) --(0,0);
			   \draw (0,1) --(2,0);
			%	\node[below] at (4.5,-0.2) {$u_2$};
				%\node[below] at (6,-0.2) {$v \equiv u_1$};

				%\draw  (0,0.05) -- (6,0.05);

			\end{tikzpicture}
			%	}
		\caption{Biconnected component.}
		
	\end{subfigure}
	\begin{subfigure}[t]{0.4\textwidth}
		\centering
		\begin{tikzpicture}
			[scale=1.2,auto=left]
			\node[style={circle,fill=black,scale=0.5}] (A) at (0,0) {};
			
			\node[style={circle,fill=black,scale=0.5}]  (B) at (1,-0.6)  {};
			
			\node[style={circle,fill=black,scale=0.5}]  (C) at (2,0) {};
			
			\node[style={circle,fill=black,scale=0.5}]  (D) at (2,1) {};
			
			\node[style={circle,fill=black,scale=0.5}]  (E) at (0,1) {};
			
				\node[style={diamond,fill=red,scale=0.7}]  (F) at (1,0.3) {};
			\draw  (0,0) -- (1,0.3);
				\draw  (B) -- (1,0.3);
					\draw  (C) -- (1,0.3);
						\draw  (D) -- (1,0.3);
			\draw  (E) -- (F);
		
			\node[style={diamond,fill=black,scale=0.7}]  (G) at (1,0.3) {};
			%	\node[below] at (4.5,-0.2) {$u_2$};
			%\node[below] at (6,-0.2) {$v \equiv u_1$};

			%\draw  (0,0.05) -- (6,0.05);

		\end{tikzpicture}
		%	}
	\caption{Star subgraph with a trans-shipment internal vertex (diamond shape). }
	
\end{subfigure}

	\vspace{5mm}
	
	\caption{Conversion of a biconnected component of a graph into a star subgraph.}
	\label{fig:tsv}

\end{figure}

The more general MAPF problem can always be reduced to a variant of the PMT (called \textit{ts}-PMT), both in the case of undirected graphs  \cite{tree,PMT} and of directed graphs \cite{diSC}. Given a graph $G$, it is possible to convert it into a tree (called \textit{biconnected component tree}), adding a new type of vertex called \textit{trans-shipment}  \cite{diSC,tree,PMT}. In particular, each biconnected component of the graph is converted into a star subgraph, 
whose internal vertex is a trans-shipment (see Figure \ref{fig:tsv}). This way, the original problem on graph $G$ is converted into a problem over a tree with trans-shipment vertices. Once a solution of the problem over the tree is obtained, this can be converted back into a solution for the original problem.

For this reason, we need to study a variant of the PMT problem, the \textit{pebble motion on trees with trans-shipment vertices} (\textit{ts}-PMT), which is a PMT problem on a tree such that the vertex set is partitioned in trans-shipment and regular vertices.
\begin{defn}
	\label{deftv}
	A \textit{trans-shipment vertex} is a vertex with degree greater than one that cannot host a pebble: pebbles can cross this node, but cannot stop there. More formally, given a trans-shipment vertex $s$,
	
	\begin{enumerate}
		\item $\text{deg}(s) \geq 2$;
		
		\item   $\rho(\A,(u \rightarrow s)(w \rightarrow v))!$ if and only if $w=s$, $(u,s),(s,v)\in E$, and $\A^{-1}(v) \in H$. If $\rho(\A,(u \rightarrow s)(s \rightarrow v))!$, then $\rho(\A,(u \rightarrow s)(s \rightarrow v))= \A[u,v]$.
	\end{enumerate}

\end{defn}
The second property  means that, if a pebble is moved to a trans-shipment vertex, then it must be immediately moved to another node.

We denote by $V_T$ the set of all the trans-shipment vertices and $V_R=V \setminus V_T$ the set of regular vertices. 
We require that $V_T$ satisfies the following property
\begin{equation}
	\label{dist}
	\forall \; v,w \in V_T \quad \quad d(v,w)>1.
	\end{equation}
 
This assumption is motivated by the fact that trans-shipment vertices are the internal vertices of the stars, so they cannot be adjacent to each other.

Now we can formally define the PMT problem with trans-shipment vertices as follows.
	\begin{defn}{(\textbf{PMT problem with trans-shipment vertices}).}
	Let $T=(V,E)$ be a tree with $V=V_R \cup V_T$, where the set of trans-shipment vertices $V_T$ is such that (\ref{dist}) holds. Given a pebble set $P$, initial and final valid configurations $\tilde{\A}^s$, $\tilde{\A}^t$ such that $\tilde{\A}^s(P),\tilde{\A}^t(P) \subset V_R$, find a plan $f$ such that $\tilde{\A}^t= \rho(\tilde{\A}^s,f)$.
\end{defn}
This problem can be solved with the same procedure described in Section \ref{sec:pmt}. However, some changes need to be made to ensure that the second property  of Definition \ref{deftv} is fulfilled. 
In the next subsections we show the changes we need to introduce into the previous procedures to address the presence of trans-shipment vertices.

\subsection{Basic plans}
\begin{figure}[h!]
	\centering
	%	\resizebox{\columnwidth}{!}{
		\begin{subfigure}[t]{0.5\textwidth}
			\centering
			\begin{tikzpicture}
				[scale=1.2,auto=left]
				\node[style={circle,fill=blue!30}] (A) at (0,0) {\textbullet};
				\node[style={rectangle,fill=green!40}] (B) at (1.5,0) {\textbullet};
				\node[below] at (0,-0.2) {$w \equiv u_5$};
				\node[style={rectangle,fill=green!40}] (J) at (1.5,1.5) {\textbullet};
				\node[below] at (1.5,-0.2) {$u_4$};
				\node[style={circle,fill=blue!30,scale=1.7}] (C) at (3,0) {};
				 \node[style={diamond,fill=black,scale=0.5}] (C1) at (3,0.05) {};
				\node[below] at (3.2,-0.2) {$u_3$};
				%	\node[below] at (A) {$s$};
				%	\node[above] at (A) {$f_0$};
				%	\node[below] at (B) {$i_0$};
				%	\node[above] at (C) {$i_1$};
				\node[style={circle,fill=blue!30}] (K) at (3,-1.5) {\textbullet};
				
				\draw  (1.5,0) -- (1.5,1.5);
				\draw  (3,0) -- (3,-1.5);
				%\node[left] at (K) {$f_1$};
				\node[style={rectangle,fill=green!40}] (D) at (4.5,0) {\textbullet};
				%	\node[above] at (D) {$z_2$};
				\node[style={rectangle,fill=green!40}] (E) at (6,0) {\textbullet};
				%	\node[below] at (6.15,0) {$s$};
				\node[below] at (4.5,-0.2) {$u_2$};
				\node[below] at (6,-0.2) {$v \equiv u_1$};

				\draw  (0,0.05) -- (6,0.05);
				
				\path[->]  (B)  edge[bend right=40,draw=red] node[above]{{\color{red}$(1)$}}  (A);
				\path[->]  (C)  edge[bend right=40,draw=red]   node[above]{{\color{red}$(3)$}} (B);
				\path[->]  (D)  edge[bend right=40,draw=red]  node[above]{{\color{red}$(2)$}}  (C);
				\path[->]  (E)  edge[bend right=40,draw=red]  node[above]{{\color{red}$(4)$}}  (D);
			\end{tikzpicture}
			\caption{Initial configuration. The red edges represent plan $\alpha_{vw}=(u_{4}\rightarrow w,u_{2}\rightarrow u_3,u_{3}\rightarrow u_4,v\rightarrow u_{2})$. Node $u_3$ (identified by the diamond shape) is a trans-shipment vertex}
			
		\end{subfigure}
		
		\vspace{5mm}
		
		\begin{subfigure}[t]{0.5\textwidth}
			\centering
			\begin{tikzpicture}
				[scale=1.2,auto=left]
				\node[style={rectangle,fill=green!40}] (A) at (0,0) {\textbullet};
				\node[style={rectangle,fill=green!40}] (B) at (1.5,0) {\textbullet};
				\node[below] at (0,-0.2) {$w\equiv u_5$};
				\node[style={rectangle,fill=green!40}] (J) at (1.5,1.5) {\textbullet};
				\node[below] at (1.5,-0.2) {$u_4$};
				\node[style={circle,fill=blue!30,scale=1.7}] (C) at (3,0) {};
			\node[style={diamond,fill=black,scale=0.5}] (C1) at (3,0.05) {};
				\node[below] at (3.2,-0.2) {$u_3$};
				%	\node[below] at (A) {$s$};
				%	\node[above] at (A) {$f_0$};
				%	\node[below] at (B) {$i_0$};
				%	\node[above] at (C) {$i_1$};
				\node[style={circle,fill=blue!30}] (K) at (3,-1.5) {\textbullet};
				
				\draw  (1.5,0) -- (1.5,1.5);
				\draw  (3,0) -- (3,-1.5);
				%\node[left] at (K) {$f_1$};
				\node[style={rectangle,fill=green!40}] (D) at (4.5,0) {\textbullet};
				%	\node[above] at (D) {$z_2$};
				\node[style={circle,fill=blue!30}] (E) at (6,0) {\textbullet};
				%	\node[below] at (6.15,0) {$s$};
				\node[below] at (4.5,-0.2) {$u_2$};
				\node[below] at (6,-0.2) {$v \equiv u_1$};

				\draw  (0,0.05) -- (6,0.05);

			\end{tikzpicture}
			%	}
		\caption{Final configuration after bringing the hole from $w$ to $v$.}
		
	\end{subfigure}
	\vspace{5mm}
	
	\caption{Example of \textsc{Bring hole from $w$ to $v$}. Vertex $u_3$ is a trans-shipment. Green squares represent pebbles, blue circles represent holes.}
	\label{bring2}

\end{figure}
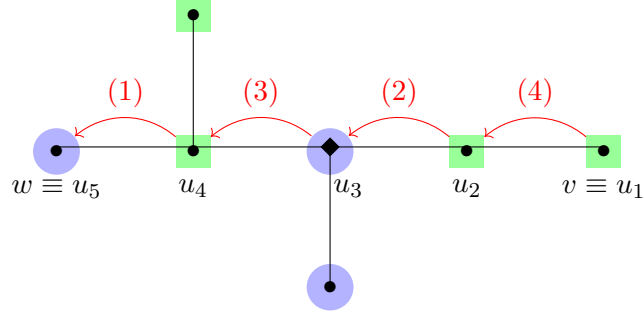
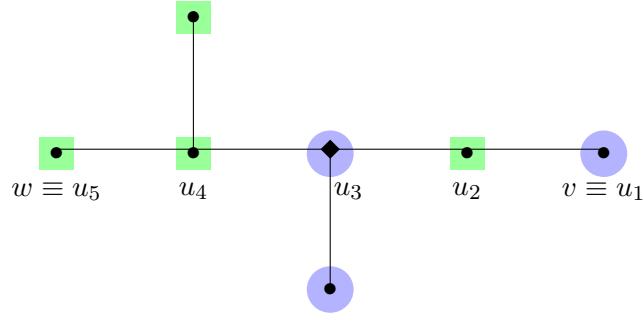

We generalize the definition of the plan \textsc{Bring hole from $w$ to $v$} to the case in which there is a trans-shipment vertex on the path $\pi_{vw}$. For instance, if $\pi_{vw}=v \, u_2\, \cdots u_i \cdots \, u_{n-1} \, u_n\equiv w$ such that $u_i \in V_T$,  then $\alpha_{vw}$ is defined as follows 	
\begin{equation}
	\label{eq:bringhole2}
	(u_{n-1}\rightarrow w,\ldots,u_{i-1}\rightarrow u_i, u_{i}\rightarrow u_{i+1}, \ldots,v\rightarrow u_{2}).
\end{equation}
In other words, 
the only difference from the previous definition is that if a pebble move on $u_i$, then it immediately moves to  $u_{i+1}$. For instance, see the example of Figure \ref{bring2}, where node $u_3$ is a trans-shipment vertex.

\begin{obser}
	\label{obts}
	Note that $\alpha_{vw}$ can be defined only if $w \in \A(H)\cap V_R$, which means that it is not allowed to bring hole from a trans-shipment vertex. Indeed, this could imply that in the final configuration a pebble lands on $w$. For the same reason plan \textsc{Move Pebble from $v$ to $w$} (i.e., $\beta_{vw}$), which in this case does not change, can be defined only if $w \in \A(H)\cap V_R$.
\end{obser}

\subsection{Assumption}
Observation \ref{obts} implies that the main difference of the new algorithm is that the holes on the trans-shipment vertices cannot be used in bring hole and gather hole operations, which are the basis for all the procedures that constitute the algorithm to solve PMT. For this reason, we define a new distance $\tilde{d}$ which does not take into account trans-shipment vertices. Given a path $\pi_{uv}$, $\tilde{d}(u,v)$ counts how many regular vertices belong to the path:

\[\tilde{d}(u,v) :=  |\pi_{uv} \cap V_R |.\]

Consequently, we also define $\tilde{c}_1$ and $\tilde{c}_2$ which count corridor lengths according to the new definition of distance:

\[\tilde{c}_1 := \max \{\tilde{d}(a,b) : \, \pi_{ab} \in C(T)\},\]
\[\tilde{c}_2 := \max \{\tilde{d}(a,b) : \, \pi_{ab} \in \bar{C}(T)\}.\]
Moreover, we define $\tilde{c}:=\tilde{c}_1$ in the case of a path graph,  $\tilde{c}:=\max \{\tilde{c}_1 +1,\, \tilde{c}_2 +2\}$ otherwise. We note that on a tree with $V_T = \emptyset$, it holds that $ d(u,v)= \tilde{d}(u,v) -1$ and $c = \tilde{c}-1$. Thus, 
to ensure the feasibility of any $ts$-PMT instance, at least $\tilde{c}-1$ holes on regular vertices  and $|V_T|$ holes for all trans-shipment vertices are needed.
Therefore, Assumption (\ref{ass}) becomes
\begin{equation}\label{ass2}
	|H| \geq |V_T| + \tilde{c}-1.
\end{equation}

\subsection{Unlabeled PMT with trans-shipment vertices}
To solve this problem, we use the same procedure described in Section \ref{sec:upmt} to solve the classical \textit{Unlabeled} PMT. 
The only difference is in Step 2 in the case vertex $v$ is a source but not a target ($v \in S \setminus D$). Here, we need an unoccupied vertex $u$ in order to move each pebble on the path $\pi_{vu}$ towards it with plan $\alpha_{vu}$. In this case $u$ must be a regular vertex, so that we need to replace (\ref{eq:choiceu}) with:

\[
u \in \arg \min_{v' \in V_R \setminus S
} d(v',v).
\]

\subsection{Gather holes problem with trans-shipment vertices}
We use the same procedure described in Section \ref{sec_gather}. However, the choice of set $M$ defined
in (\ref{eq:choiceM}) needs to be replaced by:
$$
M\in \arg\min_{W\subset \A^s(H) \cap V_R:\ |W|=q} d(W,\bar{V}),
$$
to guarantee that the holes in $M$ are at regular vertices.
\subsection{Motion planning problem with trans-shipment vertices}
\begin{figure}[h!]
	\centering
	\resizebox{\columnwidth}{!}{
		\begin{tikzpicture}
			\node(A) at (0,0) {\textbullet};
		%	\node (B) at (1.5,0) {\textbullet};
				\node[style={diamond,fill=black,scale=0.6}] (B) at (1.5,0.05) {};
			\node (J) at (1.5,1.5) {\textbullet};
			\node (C) at (3,0) {\textbullet};
			\node[below] at (A) {$r$};
			\node[above] at (A) {$\ell_0$};
			\node[below] at (B) {$i_0$};
			\node[above] at (C) {$i_1$};
			\node[style={diamond,fill=black,scale=0.6}] (K) at (3,-1.5) {};
				\node (K1) at (3,-3) {\textbullet};
					\node (K2) at (4.5,-1.55) {\textbullet};
			\draw  (1.5,0) -- (1.5,1.5);
			\draw  (3,0) -- (3,-3);
			\draw  (3,-1.5) -- (4.5,-1.5);
			\node[right] at (K2) {$\ell_1$};
			\node(D) at (4.5,0) {\textbullet};
			
			\node (E) at (6,0) {\textbullet};
			\node[above] at (E) {$j_0$};
			\node (F)[style={diamond,fill=black,scale=0.6}] at (7.5,0.05) {} ;
			\node (L) at (7.5,-1.5) {\textbullet};
			\node[left] at (L) {$\ell_2$};
			\draw  (7.5,0) -- (7.5,-1.5);
			\node[above] at (F) {$j_1=i_2$};
			\node (G) at (9,0) {\textbullet} ;
			\draw  (9,0) -- (9,1.5);
			\node (M) at (9,1.5) {\textbullet};
			\node (H) at (10.5,0) {\textbullet} ;
			\node (I) at (12,0) {\textbullet} ;
			\node[above] at (I) {$t=j_2$};
			%	\draw  (12,0) -- (12,-1.5);
			%	\node (N) at (12,-1.5) {\textbullet};
			\draw  (0,0.05) -- (12,0.05);
			
			\node (O) at (0.2,1) {{\color{red}$S_0$}};
			\node (O2) at (6,1) {{\color{green}$S_1$}};
			\node (O) at (11,1) {{\color{blue}$S_2$}};
			
			\draw[draw = red] (-0.2,0.5) -- (6.3,0.5);
			\draw[draw = red] (-0.2,-2) -- (6.3,-2);
			\draw[draw = red] (-0.2,-2) -- (-0.2,0.5);
			\draw[draw = red] (6.3,0.5) -- (6.3,-2);
			
			\draw[draw = green] (2.5,0.8) -- (8.1,0.8);
			\draw[draw = green] (2.5,-1.7) -- (8.1,-1.7);
			\draw[draw = green] (2.5,0.8) -- (2.5,-1.7);
			\draw[draw = green] (8.1,0.8) -- (8.1,-1.7);
			
			\draw[draw = blue] (6.7,0.5) -- (12.5,0.5);
			\draw[draw = blue] (6.7,-2) -- (12.5,-2);
			\draw[draw = blue] (6.7,0.5) -- (6.7,-2);
			\draw[draw = blue] (12.5,0.5) -- (12.5,-2);
			%\path [->] (A)  edge[bend right=60]  node[above] {$0$} (D);
			%\path [->] (A)  edge[bend right=80]  node[above] {$0$} (E);

			%\path [->] (B)  edge[bend left=80]  node[above] {$\beta_1$} (E);

			%	\path  (F) edge  (I);
			%	\path [->] (E)  edge[bend left=40]   (H);
			%	\path [->] (E)  edge[bend right=40]   (G);
			%	\path [->] (E) edge  (F);
			%	\path [->] (G)  edge[bend right=40]   (I);
			%	\path [->] (H)  edge[bend left=40]   (I);

		\end{tikzpicture}
	}
	\caption{We consider the motion planning problem with source vertex $r$ and target vertex $t$ on a tree with $\tilde{c}=5$. 
		Diamond shapes represent trans-shipment vertices. $S_0$, $S_1$ and $S_2$ are the \textit{caterpillar sets} along path $\pi_{rt}$.}
	\label{caterpillar2}
      \end{figure}
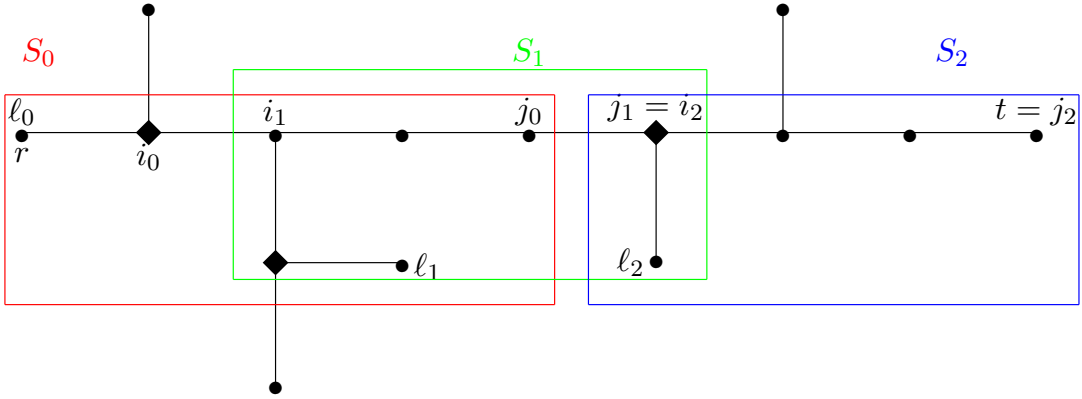

We must take into account the fact that trans-shipment vertices cannot host the marked pebble or the obstacles. Therefore, 
to ensure that the obstacle moves are feasible, we cannot only consider the cardinality of caterpillar sets, but the number of regular vertices they contain. 
To ensure this, we have to introduce the following modifications in the construction of the caterpillar sets:

\begin{enumerate}
	
	\item replace $d$ and $c$ with $\tilde{d}$ and $\tilde{c}$;
	
	\item the request on the size of the caterpillar sets concerns only the regular nodes: $|S_k \cap V_R|= \tilde{c}$ for all $k=0,\dots,m-1$, and $|S_m \cap V_R| \leq \tilde{c}$;

	\item parking positions $\ell_k$ cannot be trans-shipment vertices. At each step $k$, if the neighbors of $i_{k}$  not belonging to $\pi_{rt}$ are all trans-shipment vertices, then let $\ell_k$ be one of the 2-hop neighbors of $i_{k}$, which certainly exist in view of the first property of Definition \ref{deftv} and  are regular vertices because of assumption (\ref{dist}). Therefore, we can generalize the definition of \textit{caterpillar sets} as follows:

	\[S_k = \pi_{i_k j_k} \cup  \pi_{i_k \ell_k}\cup \pi_{i_{k+1}\ell_{k+1}}, \quad \forall k=0,\dots,m-1,\]
	
	\[S_m = \pi_{i_m j_m} \cup \pi_{i_m \ell_m}.\]
For instance, see Figure \ref{caterpillar2}.
	
\end{enumerate}

To solve the motion planning problem, we use \textit{Procedure A} and \textit{Procedure B} with some small tweaks:
\begin{enumerate}
	\item In \textit{Procedure A}:  when we slide the obstacles, we move them from $(S_{k+1} \setminus S_{k}) \cap V_R$ to $(S_{k} \setminus S_{k+1}) \cap V_R $. Indeed, we cannot bring holes from trans-shipment vertices.
	\item In \textit{Procedure B}: at each iteration we gather the holes that are on $V_j \cap V_R$ in  $H_j$, which is a subset of $V_j \cap V_R$ of cardinality $q_j$  closest to $s$:
	
	\[
	H_j \in \arg \min_{\begin{array}{c} \scriptstyle W \in \PP(V_j \cap V_R) \\ 	
			\scriptstyle 	|W| = q_j\end{array}
	} d(W,\{s\}),
	\]
	
	where $q_j = |\A^s(H) \cap V_j \cap V_R|$.
\end{enumerate}

\section{Experimental Results}

We performed two distinct set of experiments, one regarding only the motion planning algorithm, one for the whole PMT algorithm. The algorithms have been implemented in {\tt Matlab}. They can be downloaded at {\tt \url{https://github.com/auroralab-unipr/PMT}}.

\subsection{Motion planning} 

\begin{figure}[h!]
    \centering
    \includegraphics[scale=0.7]{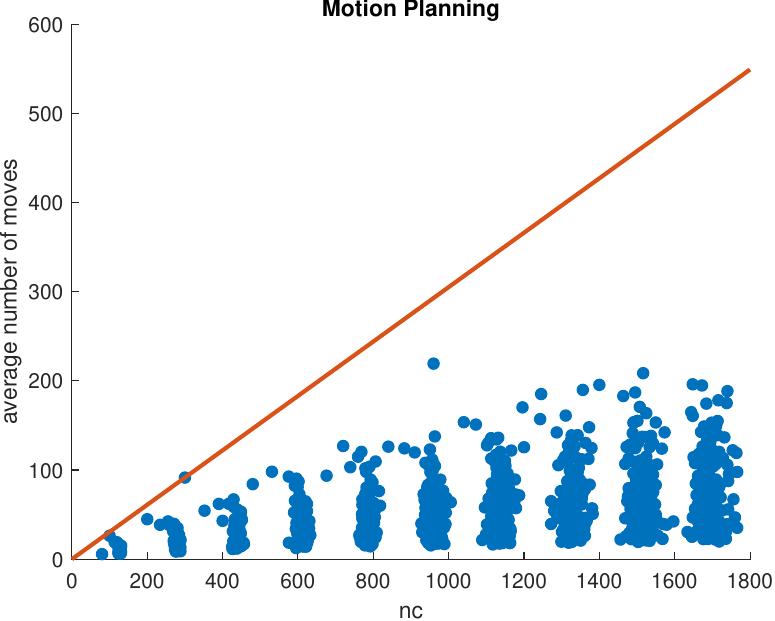}
    \caption{Average number of moves for the algorithm motion planning on $nc$.}
    \label{fig:motion}
\end{figure}

In the first set of experiments, we generated random trees with a number of nodes $|V|$ ranging from $20$ to $200$ by $20$ using the \textit{NetworkX} \cite{nx} graph generator for random trees (function \texttt{random\_tree()}). The number of agents $|P|$ ranges from $2$ to $|V|-2$, while $\A^s$ and $\A^t$ are randomly generated. Only instances that fulfill Assumption (\ref{ass}) are taken into account. For every combination of number of nodes and number of agents, we generated $100$ instances, each instance refers to a different graph. In Figure \ref{fig:motion} we display the average number of moves of the solutions found on $nc$.      

According to Propositions \ref{prop:procA} and \ref{prop:procB}, the motion planning algorithm returns solutions with length complexity $O(nc)$. In Figure \ref{fig:motion} we can see a linear upper bound for the average number of moves, that supports the complexity result. We can also see how the number of moves is often much lower than the upper bound found. We remark that instances with number of moves closer to the upper bound line are those for which the number of pebbles is very high (as expected, these are more tricky instances).

%\begin{figure}[h!]
%    \centering
%    \includegraphics[scale=0.7]{img/average_motion_planning.pdf}
%    \caption{Average number of moves for the algorithm Motion Planning on %$|P|c$.}
%    \label{fig:motion}
%\end{figure}

\subsection{PMT}

In the second set of experiments, we generated random trees with a number of nodes $|V|$ ranging from 20 to 200 by 20 using the same procedure used for the first set of experiments. The number of agents $|P|$ ranges from $5$ to $(3/4)|V|$ by $5$, while $\A^s$ and $\A^t$ are randomly generated. As for the first set of experiments, only instances that fulfill Assumption (\ref{ass}) are taken into account. For every combination of number of nodes and number of agents, we generated $20$ instances. In Figure \ref{fig:pmt} we display the average number of moves of the solutions found on $n|P|c+n^2$.      
\begin{figure}[h!]
    \centering
    \includegraphics[scale=0.7]{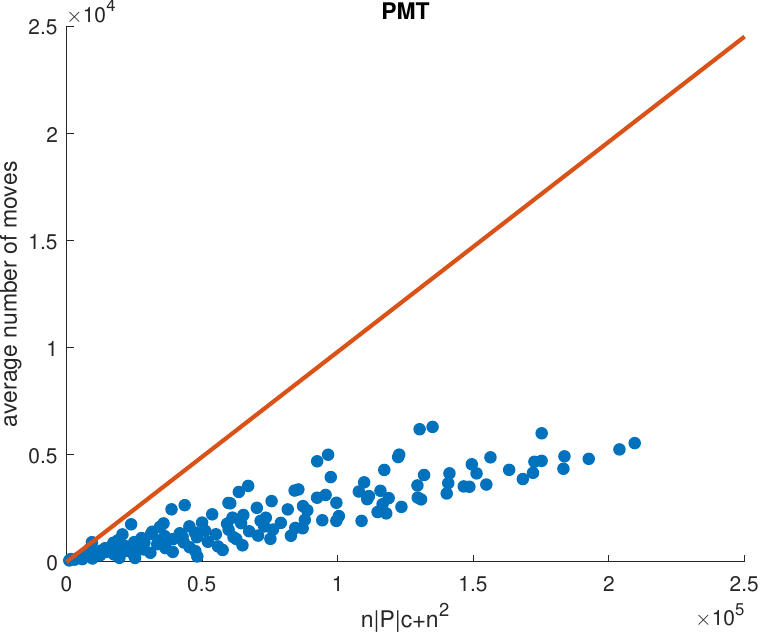}
    \caption{Average number of moves for the algorithm PMT on $n|P|c+n^2$.}
    \label{fig:pmt}
\end{figure}

As stated in Theorem \ref{th:pmt_comp}, the length complexity of the PMT algorithm is $O(n|P|c+n^2)$. Figure \ref{fig:pmt} displays a linear upper bound on the average number of moves, therefore confirming the complexity result. 

%\begin{figure}[h!]
%    \centering
%    \includegraphics[scale=0.7]{img/average_pmt.pdf}
%    \caption{Average number of moves for the algorithm PMT on $n|P|c$.}
%    \label{fig:pmt}
%\end{figure}
    \section{  Conclusion}
   
 In this paper we proposed two algorithms with improved length complexity for the motion planning problem and the pebble motion problem on trees. Denoting by $n$ the number of nodes, $c$ the maximum length of corridors and $k$ the number of pebbles,  the CATERPILLAR algorithm solves the motion planning problem with $O(nc)$ moves, while the \textit{Leaves procedure} solves the PMT problem in $O(knc + n^2)$ moves. 
 
 Moreover, we discuss a variant of the PMT problem, the PMT with trans-shipment vertices ($ts$-PMT), which considers a new type of vertex that cannot host pebbles. This problem is very interesting since MAPF instances on graphs can be reduced to it, and we proved that it can be solved with the \textit{Leaves procedure for} PMT with some minor modifications.
 
As a topic for future research, we will study pebble motion, also known as Multi Agent Path Finding (MAPF), on general graphs. As already mentioned (see Section \ref{tspmt}), the solution of MAPF on a general graph can be obtained by first converting it on the trans-shipment variant of PMT, and then converting back the obtained solution over the general graph.  An upper bound for the solution length can be derived by exploiting the complexity results of this paper.

%\bliography{references,caterpillarlio}
\vskip 0.2in
%\bliography{caterpillar}
%\bliographystyle{plain}
\printbibliography

\end{document}